\documentclass[12pt, a4paper]{article}
\pdfoutput=1
\usepackage{graphicx}
\usepackage{amssymb}
\usepackage{amsmath}
\usepackage{bm}
\usepackage[table,dvipsnames]{xcolor} 
\usepackage{cite}
\usepackage{slashed}
\usepackage{epstopdf}            
\usepackage{epsfig}
\usepackage{here}
\usepackage{comment}
\usepackage{booktabs} 
\usepackage{colortbl} 
\usepackage{wrapfig}
\usepackage{ascmac}
\usepackage{fancybox}  
\usepackage{soul} 
\usepackage{adjustbox}
\usepackage{multirow}
\usepackage{array}
\usepackage{arydshln}
\usepackage{here}

\allowdisplaybreaks
\renewcommand{\thefootnote}{\fnsymbol{footnote}}
\numberwithin{equation}{section} 
\def\beq#1\eeq{\begin{align}#1\end{align}}
\newcommand{\ov}{\overline}
\newcommand{\eg}{\textit{e.g.}}
\newcommand{\ie}{\textit{i.e.}}

\newcommand{\Lb}{\ell}%
\RequirePackage{xspace}
\usepackage{caption}
\definecolor{BlueViolet}{rgb}{0.2, 0.00, 0.7}
\definecolor{Blue}{rgb}{0.15, 0.00, 0.9}
\definecolor{light_blue}{rgb}{0.15, 0.35, 0.95}
\definecolor{kit_green}{rgb}{0
, 0.58823 
, 0.50980 
}
\usepackage[
colorlinks=true, linkcolor=light_blue,citecolor=light_blue,urlcolor=kit_green]{hyperref} 

\begin{document}
\sloppy 
\begin{titlepage}
\begin{center}
\hfill{KEK--TH--2828}\\
\vskip .3in

{\Large{\bf $b \to c$ semileptonic
sum rule:\\ \vskip .04in
orbitally excited hadrons}}

\vskip .3in

\makeatletter\g@addto@macro\bfseries{\boldmath}\makeatother

{ 
Motoi Endo$^{\rm (a,b)}$,\,
Syuhei Iguro$^{\rm (c,d)}$,\,
Satoshi Mishima$^{\rm (e)}$
}
\vskip .3in
$^{\rm (a)}${\it KEK Theory Center, IPNS, KEK, Tsukuba 305--0801, Japan}\\\vspace{4pt}
$^{\rm (b)}${\it Graduate Institute for Advanced Studies, SOKENDAI, Tsukuba,\\ Ibaraki 305--0801, Japan} \\\vspace{4pt}
$^{\rm (c)}${\it Institute for Advanced Research (IAR), Nagoya University,\\ Nagoya 464--8601, Japan}\\\vspace{4pt}
$^{\rm (d)}${\it Kobayashi-Maskawa Institute (KMI) for the Origin of Particles and the Universe, Nagoya University, Nagoya 464--8602, Japan}\\\vspace{4pt}
$^{\rm (e)}${\it Department of Liberal Arts, Saitama Medical University, Moroyama,\\ Saitama 350--0495, Japan}
\end{center}
\vskip .15in

\begin{abstract}
We study semileptonic sum rules for $b \to c \tau \overline{\nu}$ transitions involving orbitally excited charm hadrons.
Starting from the amplitude-level relation implied by the heavy quark symmetry, we construct sum rules relating these decays.
We then examine deviations from the small-velocity limit.
Our numerical analysis shows that the deviations generally increase once excited hadrons are involved, with tensor contributions often inducing sizable effects.
At present, however, the relevant form factors are not yet sufficiently constrained.
Further improvements in the hadronic inputs are essential for these sum rules to yield robust predictions for the corresponding lepton-universality ratios.\\
\end{abstract}
{\sc ~~~~ Keywords: $b \to c$ semileptonic sum rule, Orbitally excited states} 
\end{titlepage}

\setcounter{page}{1}
\renewcommand{\thefootnote}{\#\arabic{footnote}}
\setcounter{footnote}{0}

\hrule
\tableofcontents
\vskip .2in
\hrule
\vskip .4in

\section{Introduction}
\label{sec:intro}

The $b\to c\tau\ov\nu$ transitions among ground-state hadrons, $B\to D\tau\ov\nu$, $B\to D^*\tau\ov\nu$, and $\Lambda_b\to\Lambda_c\tau\ov\nu$, provide one of the cleanest probes of experimental data and theoretical predictions in semileptonic weak decays.
A particularly useful consequence of heavy-quark symmetry is that, in the small-velocity (Shifman–Voloshin, SV) limit, the corresponding decay rates are not independent.
In this limit, where the heavy-quark and zero-recoil limits are taken simultaneously, they satisfy a relation among the lepton-universality ratios, $R_D$, $R_{D^*}$, and $R_{\Lambda_c}$, defined as $R_{H_c}={\rm{BR}}(H_b\to H_c\tau\ov\nu)/{\rm{BR}}(H_b\to H_c \ell\ov\nu)$ with $\ell=e,\mu$. 
This relation is known as the $b \to c$ semileptonic sum rule~\cite{Endo:2025fke, Endo:2025lvy, Endo:2025set},\footnote{
Such a relation has also been discussed in the literature \cite{Blanke:2018yud,Blanke:2019qrx,Fedele:2022iib,Duan:2024ayo,Iguro:2026xgi} and, for angular observables, in Ref.~\cite{Endo:2025cvu}.
}
\begin{align}
\label{eq:3mode_2508}
    \frac{R_{\Lambda_c}}{R_{\Lambda_c}^{\rm SM}}-\frac{1}{4}\frac{R_{D}}{R_{D}^{\rm SM}}-\frac{3}{4}\frac{R_{D^*}}{R_{D^*}^{\rm SM}}=\delta\, .
\end{align}
This relation holds even in the presence of new physics (NP) contributions. 
The coefficients $1/4$ and $3/4$, together with $\delta = 0$, are fixed in the SV limit~\cite{Endo:2025lvy}.
For realistic hadrons, however, the SV limit is not exact, and $\delta$ therefore becomes nonzero.
It encodes corrections via hadron masses and hadronic form factors, as well as possible NP effects.
If $\delta$ is small compared with the experimental uncertainties, the measured values are expected to satisfy the above relation irrespective of NP contributions.

Given the usefulness of this relation for transitions between ground-state hadrons, it is natural to ask whether a similar strategy can be extended to semileptonic decays into orbitally excited charmed hadrons, such as $\Lambda_c^*$ and $D^{**}$, for example $D^{*}_2$ and $D_1$.
In this case, however, the situation is less straightforward when one attempts to construct analogous sum rules.
The reason is that the behavior of the hadronic form factors in the zero-recoil limit differs substantially from that in transitions between ground-state hadrons.
Moreover, decays into excited states are expected to be more sensitive to realistic hadron-mass effects as well as to higher-order corrections to the form factors.
The central question is therefore not only whether such sum rules can be constructed, but also whether they retain quantitative predictive power.

In this paper, we study decays of ground-state bottom hadrons into excited charm hadrons.
We first derive sum rules among decays into excited charm hadrons in the SV limit and then quantify violations of this limit for realistic hadrons.
However, relations that mix decays into ground-state charm hadrons with those into excited charm hadrons require a separate treatment, because the zero-recoil behavior of the relevant form factors differs between ground-state and excited-state final hadrons.
These cases are therefore discussed separately in the appendix.

At present, the relative experimental precision of $R_{H_c}$ has reached the few-percent level for semileptonic transitions between ground-state mesons, namely $R_D$ and $R_{D^*}$ \cite{HFLAV:2024ctg}, while it remains at the level of $\sim 30\%$ for the baryonic mode $R_{\Lambda_c}$ \cite{LHCb:2022piu}.
By contrast, for decays into excited mesons, $R_{D_2^{*}}$ and $R_{D_1}$ have been measured with a precision of $\sim 30\%$ \cite{LHCb:2025fri}.\footnote{
Among decays into $P$-wave excited hadrons, $R_{D_{0}^{*}}$, $R_{D_{1}^{*}}$, $R_{\Lambda_c^*(1/2^-)}$, and $R_{\Lambda_c^*(3/2^-)}$ have not yet been measured, although the future prospects are promising.
See Table~\ref{tab:prospect} for a summary of the experimental status.
}
In light of the expected experimental progress, we not only derive the corresponding sum rules but also examine, using the currently available form factor inputs, whether these relations can become quantitatively useful in future measurements.

The rest of this paper is organized as follows.
In section~\ref{sec:Formulation}, we introduce the effective operators describing possible NP contributions.
Section~\ref{sec:Sum_rule} is devoted to the construction of the sum rules, while section~\ref{sec:Numerical} presents the numerical results.
Finally, section~\ref{sec:Conclusion} contains our conclusions.
Sum rule coefficients and supplementary analysis are collected in the appendix.

\section{New physics framework}
\label{sec:Formulation}

In this paper, we construct sum rules in the presence of NP.
Since we consider its contributions affecting only the $b\to c \tau\ov\nu$ transitions, the weak effective Hamiltonian is generally given by 
\begin{align}
 \label{eq:Hamiltonian}
 {\cal {H}}_{\rm{eff}}= 2 \sqrt2 \, G_FV_{cb}\biggl[ (1+C_{V_L})O_{V_L}+C_{S_L}O_{S_L}+C_{S_
R}O_{S_R}+C_{T}O_{T}\biggl]\,.
\end{align}
The dimension-six operators are defined as
\begin{align}
 &O_{V_L} = (\overline{c} \gamma^\mu P_Lb)(\overline{\tau} \gamma_\mu P_L \nu_{\tau})\,,\,\,\,\,\,\,\,\,\, 
 O_{S_L} = (\overline{c} P_L b)(\overline{\tau} P_L \nu_{\tau})\,, \label{eq:operator}\nonumber \\
 &O_{S_R} = (\overline{c}  P_Rb)(\overline{\tau} P_L \nu_{\tau})\,,\,\,\,\,\,\,\,\,\,
 O_{T} = (\overline{c}  \sigma^{\mu\nu}P_Lb)(\overline{\tau} \sigma_{\mu\nu} P_L \nu_{\tau}) \,,
\end{align}
where $P_{L(R)}=(1\mp\gamma_5)/2$ denotes the chirality projection operator. 
The NP effects are encoded in the Wilson coefficients $C_X$, which are normalized to the standard model (SM) prefactor $2\sqrt{2}\, G_F V_{cb}$.
The coefficients are evaluated at the scale $\mu_b=4.2\,$GeV. 
In this framework, the SM limit corresponds to $C_{X} = 0$ for $X=V_{L}$, $S_{L,R}$, and $T$. 
We further assume that the neutrinos in Eq.~\eqref{eq:Hamiltonian} are left-handed.\footnote{
The inclusion of a massive right-handed neutrino leads to only a small violation of the sum rule for ground-state transitions \cite{Iguro-Kretz:2026}.
See also Refs.~\cite{Iguro:2018qzf,Robinson:2018gza,Babu:2018vrl, Mandal:2020htr,Penalva:2021wye,Datta:2022czw} for explicit models and analysis.
This conclusion does not necessarily extend to decays into excited states, whose impact will be investigated elsewhere.
}

\section{Sum rule involving excited hadrons}
\label{sec:Sum_rule}

\begin{table}[t]
  \centering
\scalebox{1.05}{
  \begin{tabular}{c|cc|c|c}
    Content 
    & $I\,(J^{P})$ & $s_{\Lb}^{P_{\Lb}}$ 
    & $b$ hadron & $c$ hadron 
    \\   \hline\hline
    $Qud$
    & $0\,(\tfrac12^{+})$ & $0^{+}$ 
    & $\Lambda_b$ 
    & $\Lambda_c$ \\ \hline
    $Q\ov q$
    & $\tfrac12\,(0^{-},1^{-})$
    & $\tfrac12^{-}$ 
    & $B^{1/2^-} = (B,\, B^*)$
    & $D^{1/2^-} = (D,\, D^*)$ 
    \\ \hline\hline
    $Qud$
    & $0\,(\tfrac12^{-},\tfrac32^{-})$ & $1^{-}$  
    & $\Lambda_b^* = (\Lambda_b^*(1/2^-),\,\Lambda_b^*(3/2^-))$
    & $\Lambda_c^* = (\Lambda_c^*(1/2^-),\,\Lambda_c^*(3/2^-))$
    \\ \hline
    \multirow{2}{*}{$Q\ov q$}
    & $\tfrac12\,(0^{+},1^{+})$
    & $\tfrac12^{+}$ 
    & $B^{1/2^+} = (B_0^*,\, B_1^*)$
    & $D^{1/2^+} = (D_0^*,\, D_1^*)$ \\ \cline{2-5}
    & $\tfrac12\,(1^{+},2^{+})$
    & $\tfrac32^{+}$ 
    & $B^{3/2^+} = (B_1,\, B_2^*)$
    & $D^{3/2^+} = (D_1,\, D_2^*)$ \\ \hline\hline
  \end{tabular}
  }
  \caption{Bottom ($Q=b$) and charm ($Q=c$) hadrons, where $q$ denotes $u$ or $d$. 
  The first column indicates the quark composition. 
  The first two rows correspond to ground-state hadrons, while the remaining rows list $P$-wave excited states.
  The $\Lambda_Q$ baryon is a singlet under the heavy-quark (spin) symmetry, whereas the other hadrons form doublets, which are represented as vectors.
  Here, $I$ and $J^P$ denote the isospin and spin-parity of the hadron, respectively, while $s_{\Lb}^{P_{\Lb}}$ denotes the spin-parity of the light degrees of freedom. 
  By convention, quarks (antiquarks) are assigned positive (negative) parity.}
  \label{tab:hadrons}
\end{table}

In this section, we derive sum rules involving excited hadrons. 
We consider transitions from ground-state bottom hadrons into excited charm hadrons, focusing in particular on $P$-wave excitations. 
The hadrons are characterized by their isospin, spin, and parity, and are further classified according to the heavy-quark symmetry, as summarized in Table~\ref{tab:hadrons}.
The experimental status and future prospects of $R_{H_c}$ measurements are summarized in Table~\ref{tab:prospect}.

\begin{table}[t]
\begin{center}
  \begin{tabular}{c|ccc|ccc} 
 Mode & $\Lambda_c$& $D$ & $D^*$ & $\Lambda_c^*$ & $D^{1/2^+}$& $D^{3/2^+}$  \\ \hline
Current &$31\%$&$6.7\%$ &$3.9\%$&$-$&$-$&$29\%$\\ 
Prospect &$5\,(2)\%$&$1.3\%$&$1.0\%$&$7\,(5)\%$&$-$& $5\,(3)\%$\\\hline 
\end{tabular}
  \caption{Summary of current and projected mid-term experimental relative uncertainties for $R_{H_c}$.
  The current (future) precisions are taken from Ref.~\cite{LHCb:2022piu} (Ref.~\cite{Bernlochner:2021vlv}) for $H_c = \Lambda_c$, Ref.~\cite{HFLAV:2024ctg} (Ref.~\cite{ATLAS:2025lrr}) for $D$ and $D^*$, Ref.~\cite{LHCb:2025fri} (Ref.~\cite{Bernlochner:2021vlv}) for $D^{3/2^+}$, and Ref.~\cite{Bernlochner:2021vlv} for projections for $\Lambda_c^*$.
  The recent result of $R_{D^{**}}$ is based on the measurements of the narrow $D^{3/2^+}$ states.
  For future projections, the values in parentheses correspond to an optimistic scenario for systematic uncertainties, while the others represent a pessimistic scenario. 
  The projections for $\Lambda_c^*$ and $D^{3/2^+}$ are obtained using the $\Lambda_c^*(3/2^-)$ and $D_1$ channels, which feature cleaner final states and higher statistics.
  The mid-term projections are based on the expected performance of Belle II and LHCb. 
  To the best of our knowledge, no projections are currently available for $D^{1/2^+}$.
 }
  \label{tab:prospect}
\end{center}   
\vspace{-.25cm}
\end{table}

\subsection{General structure of sum rules}
\label{sec:SR_general}

We extend the sum rule in Eq.~\eqref{eq:3mode_2508}, originally formulated for $B$ and $\Lambda_b$ decays into ground-state charm hadrons, to transitions involving orbitally excited charm hadrons.
We consider sum rules that relate the two decays $H_b \to H_c \tau\ov\nu$ and $H_b' \to H_c' \tau\ov\nu$, where $H_b$ and $H_b'$ are either $B$ or $\Lambda_b$, whereas $H_c$ and $H_c'$ are charm-hadron multiplets.
Each multiplet is either a heavy-quark singlet or doublet, as listed in Table~\ref{tab:hadrons}.
For a heavy-quark doublet, we label its two members by $(1)$ and $(2)$, for example, $H_{c}(1)=D$ and $H_{c}(2)=D^*$ for $H_c = D^{1/2^-} = (D, D^*)$. 

When both $H_c$ and $H_c'$ are heavy-quark doublets, the sum rules can be written generally in the form,
\begin{align}
 \left[
 \frac{R_{H_c(1)}}{R_{H_c(1)}^{\,\mathrm{SM}}} 
 +
 \gamma_{1}^{X} 
 \frac{R_{H_c(2)}}{R_{H_c(2)}^{\,\mathrm{SM}}} 
 \right]
 - 
 \left[
 \gamma_{2}^{X}
 \frac{R_{H_c'(1)}}{R_{H_c'(1)}^{\,\mathrm{SM}}} 
 +
 \gamma_{3}^{X}
 \frac{R_{H_c'(2)}}{R_{H_c'(2)}^{\,\mathrm{SM}}}
 \right]
 =
 \delta^{X}[H_c,H_c']
 \,.
 \label{eq:4SR_general}
\end{align}
Here the superscript $X$ is a label to specify the prescription to construct the relation introduced below.
When $H_c$ is a heavy-quark singlet and $H_c'$ a heavy-quark doublet, the sum rule takes the form 
\begin{align}
 \frac{R_{H_c}}{R_{H_c}^{\,\mathrm{SM}}} 
 - 
 \left[
 \gamma_{2}^{X}
 \frac{R_{H_c'(1)}}{R_{H_c'(1)}^{\,\mathrm{SM}}} 
 +
 \gamma_{3}^{X}
 \frac{R_{H_c'(2)}}{R_{H_c'(2)}^{\,\mathrm{SM}}} 
 \right]
 =
 \delta^{X}[H_c,H_c']
 \,.
 \label{eq:3SR_general}
\end{align}
We refer to $\delta^{X}[H_c,H_c']$ as the deviation of the sum rule. 
When its magnitude is sufficiently small, the sum rule becomes predictive. 
In that case, one of the ratios $R_{H_c}$ can be inferred from the others independently of the details of the NP contributions, as discussed later.
The corresponding measurements can therefore be subjected to a non-trivial consistency test.

Since NP is assumed to affect only the $b \to c \tau \ov\nu$ transition, $R_{H_c}$ can be generally written as
\begin{align}
 R_{H_c} 
 = 
 \sum_{i,j} \mathcal{C}_i\, \mathcal{C}_j^*\,
 R_{H_c}^{\,ij}\,,
\end{align}
where the coefficients $\mathcal{C}_i$ are defined by $\mathcal{C}_{V_L} = 1+C_{V_L}$ and $\mathcal{C}_i=C_i$ for $i = S_L, S_R, T$. 
The SM prediction corresponds to $C_{X} = 0$, so that $R_{H_c}^{\,\mathrm{SM}} = R_{H_c}^{V_L V_L}$.
Since $R_{H_c}^{S_L S_L} = R_{H_c}^{S_R S_R}$ holds in general, we denote both $(ij) = (S_L S_L)$ and $(S_R S_R)$ collectively as $(ij) = (SS)$. 
Accordingly, the deviation $\delta^{X}[H_c,H_c']$ is decomposed as
\begin{align}
 \delta^{X}[H_c,H_c'] 
 = 
 \sum_{i,j} \mathcal{C}_i\, \mathcal{C}_j^* \,
 \delta^{X}_{ij}[H_c,H_c']
 \,, 
 \label{eq:deltaX}
\end{align}
where $\delta^{X}_{ij}[H_c,H_c']$ is defined by replacing $R_{H_c}$ with $R_{H_c}^{ij}$ in the numerators on the right-hand sides of Eqs.~\eqref{eq:4SR_general} and \eqref{eq:3SR_general}. 
The numerical expressions for $R_{H_c}$ are summarized in Appendix~\ref{app:fitting_formula}.

In this work, we consider two prescriptions of fixing the coefficients $\gamma_{i}$. 
One is based on the SV limit, while the other follows the approach developed by the KIT group~\cite{Blanke:2018yud, Blanke:2019qrx, Fedele:2022iib}, which we refer to as the KIT prescription.
In both cases, the coefficients are constrained by the requirement that $\delta^{X}[H_c,H_c']$ vanishes in the SM limit, $R_{H_c} \to R_{H_c}^{\,\mathrm{SM}}$.
This implies
\begin{align}
1 + \gamma_{1}^{X} = \gamma_{2}^{X} + \gamma_{3}^{X}
\label{eq:gamma_constraint_4}
\end{align}
for Eq.~\eqref{eq:4SR_general}, and 
\begin{align}
1 = \gamma_{2}^{X} + \gamma_{3}^{X}
\label{eq:gamma_constraint_3}
\end{align}
for Eq.~\eqref{eq:3SR_general}. 
These conditions are equivalent to $\delta_{V_L V_L}^{X}[H_c, H_c'] = 0$, \ie, to requiring that the purely SM contribution to the deviation vanish.
The remaining freedom is fixed by an additional condition on the $\gamma_{i}^{X}$, which defines the prescription labeled by $X$.

\subsection{SV-limit prescription}
\label{sec:SR_SV}

The SV-limit prescription is based on a relation among the decay amplitudes in the heavy-quark limit~\cite{Endo:2025fke, Endo:2025set}.
In this limit, heavy-quark symmetry allows the hadronic amplitudes to be matched onto a universal partonic amplitude as\footnote{
The decay rate formulae for $B\to D^{(*)}\tau\ov\nu$ and $\Lambda_b\to\Lambda_c \tau\ov\nu$ are summarized, for instance, in Refs.~\cite{Endo:2025fke, Bernlochner:2018bfn}.
For decays into excited hadrons, the corresponding formulae are given in Ref.~\cite{Bernlochner:2017jxt} for $B\to D^{**}\tau\ov\nu$ and in Ref.~\cite{Papucci:2021pmj} for $\Lambda_b\to\Lambda_c^*\tau\ov\nu$.
}
\begin{align}
\label{eq:amplitude}
\frac{1}{2}
\sum_{\lambda_c,\lambda_b} 
\big| \mathcal{M}_{\lambda_c,\lambda_b} \big|^2 &= 
\Delta(\Lambda_b\to \Lambda_c\tau\ov\nu) =
\Delta(B\to D^{1/2^-}\tau\ov\nu) \nonumber \\
&=
\Delta(B\to D^{1/2^+}\tau\ov\nu)
=
\Delta(B\to D^{3/2^+}\tau\ov\nu)
=
\Delta(\Lambda_b\to \Lambda_c^*\tau\ov\nu)
\,.
\end{align}
Here, $\mathcal{M}_{\lambda_c, \lambda_b}$ denotes the partonic decay amplitude for $b \to c \tau \ov\nu$, where $\lambda_Q$ labels the polarization of the heavy quark $Q$.
This relation holds both in the SM and in the presence of NP contributions described by Eq.~\eqref{eq:Hamiltonian}. 
The first line corresponds to the relation for decays into ground-state hadrons and reproduces the result of Ref.~\cite{Endo:2025fke}.
Since the $D$ and $D^{*}$ mesons form a heavy-quark doublet $D^{1/2^-}=(D,\,D^*)$, the decay $B\to D^{1/2^-}\tau\ov\nu$ includes both $B\to D\tau\ov\nu$ and $B\to D^*\tau\ov\nu$. 
The remaining relations involve decays into excited hadrons and, except for the last one, coincide with those reported in Ref.~\cite{Endo:2025set}.

The quantities $\Delta(H_b \to H_c\tau\ov\nu)$ are defined in terms of the hadronic decay amplitudes as
\begin{align}
\Delta(H_b\to H_c\tau\ov\nu) =
\frac{
\displaystyle
\sum_{J_{H_c}}
\frac{1}{2J_{H_b}+1}
\sum_{\lambda_{H_c},\lambda_{H_b}}\!\!\!
\big|\mathcal{M}_{H_c}^{\lambda_{H_c},\lambda_{H_b}}\big|^2
}{\phantom{\Bigg|}
\displaystyle
\frac{1}{2s_{\ell}+1}
\sum_{\lambda_{\ell}} 
\big| \langle L_{H_c} | L_{H_b} \rangle \big|^2
}
\,,
\label{eq:Delta}
\end{align}
where $J_{H_Q}$ and $s_\ell$ denote the total angular momenta of the hadron and the light degrees of freedom, respectively, while $\lambda_{H_Q}$ and $\lambda_\ell$ are their corresponding projections.
They are explicitly given by
\begin{align}
&\Delta(B\to D^{1/2^-}\tau\ov\nu)
=
\frac{
\displaystyle
\big| \mathcal{M}_{D} \big|^2
+
\sum_{\lambda_{D^*}} 
\big| \mathcal{M}_{D^*}^{\lambda_{D^*}}\big|^2
}{
\langle\ov{q} (\tfrac{1}{2}^{-}, \tfrac{1}{2})|\ov{q} (\tfrac{1}{2}^{-}, \tfrac{1}{2})\rangle^2
}\,, \\
&\Delta(B\to D^{1/2^+}\tau\ov\nu)
=
\frac{\displaystyle
\sum_{H_c=D_0^*,D_1^*}
\sum_{\lambda_{H_c}}
\big|
  \mathcal{M}_{H_c}^{\lambda_{H_c}}
\big|^2
\vphantom{\bigg|}
}{\langle
  {\ov{q} (\tfrac{1}{2}^{+}, \tfrac{1}{2})}|
  {\ov{q} (\tfrac{1}{2}^{-}, \tfrac{1}{2})}
  \rangle^2
}\,, \\
&\Delta(B\to D^{3/2^+}\tau\ov\nu)
=
\frac{\displaystyle
\sum_{H_c=D_1,D_2^*}
\sum_{\lambda_{H_c}}
\big|
  \mathcal{M}_{H_c}^{\lambda_{H_c}}
\big|^2
\vphantom{\bigg|}
}{\langle
  {\ov{q} (\tfrac{3}{2}^{+}, \tfrac{1}{2})}|
  {\ov{q} (\tfrac{1}{2}^{-}, \tfrac{1}{2})} 
 \rangle^2
}\,, \\
&\Delta(\Lambda_b\to \Lambda_c^*\tau\ov\nu)
=\frac{\displaystyle
\frac{1}{2} \sum_{H_c=\Lambda_c^*(1/2^-),\Lambda_c^*(3/2^-)}
\,\sum_{\lambda_{H_c},\lambda_{\Lambda_b}}
\big|
  \mathcal{M}_{H_c}^{\lambda_{H_c},\lambda_{\Lambda_b}}
\big|^2
\vphantom{\bigg|}
}{\langle
  {qq (1^{-}, 0)}|
  {qq (0^{+}, 0)}
  \rangle^2
}\,.
\end{align}
The denominators represent matrix elements among the light degrees of freedom $L_{H_Q}$ inside the hadrons $H_Q$.
Their wave functions $|L_{H_Q}(s_\ell^{P_\ell}, \lambda_\ell) \rangle$ are characterized by their total angular momentum $s_\ell$, its projection $\lambda_\ell$, and parity $P_\ell$. 
The matrix elements, describing their overlaps, are functions of the kinematic variable $w = (m_{H_b}^2 + m_{H_c}^2 - q^2)/(2 m_{H_b} m_{H_c})$, where $q^2$ is the invariant mass of the lepton pair.

In the heavy-quark limit, these matrix elements can be expressed in terms of the leading-order (LO) Isgur–Wise (IW) functions~\cite{Sadzikowski:1993iv}.
For transitions between ground-state hadrons, the LO IW functions for mesons and baryons~\cite{Isgur:1989vq, Isgur:1990yhj, Isgur:1990pm, Georgi:1990cx} are given by 
\begin{align}
\xi(w)
=
\sqrt{\frac{2}{w+1}}\,
\langle
  {\ov{q} (\tfrac{1}{2}^{-}, \tfrac{1}{2})}|
  {\ov{q} (\tfrac{1}{2}^{-}, \tfrac{1}{2})}
\rangle\,,
~~~
\zeta(w)
=
\langle
  {qq (0^{+},0)}|{qq (0^{+},0)}
\rangle
\label{eq:zeta}
\,.
\end{align}
Due to heavy-quark symmetry, transitions into members of the same heavy-quark multiplet, \eg, the $B\to D$ and $B\to D^*$ transitions, are governed by a common IW function.
For transitions into the $P$-wave excited meson doublets, the corresponding IW functions $\tau_{1/2}(w)$ and $\tau_{3/2}(w)$~\cite{Isgur:1990jf, Falk:1992wt, Leibovich:1997em, Leibovich:1997tu} are given by
\begin{align}
\tau_{1/2}(w)
&=
\frac{1}{\sqrt{2(w-1)}}\,
\langle
  {\ov{q} (\tfrac{1}{2}^{+}, \tfrac{1}{2})}|
  {\ov{q} (\tfrac{1}{2}^{-}, \tfrac{1}{2})}
\rangle
\,,
\label{eq:IW_tau12}
\\
\tau_{3/2}(w)
&=
-
\frac{1}{(w+1)}\frac{1}{\sqrt{w-1}}\,
\langle
  {\ov{q} (\tfrac{3}{2}^{+}, \tfrac{1}{2})}|
  {\ov{q} (\tfrac{1}{2}^{-}, \tfrac{1}{2})}
\rangle
\,.
\label{eq:IW_tau32}
\end{align}
For transitions into the negative-parity excited baryon doublet, the IW function $\sigma(w)$~\cite{Isgur:1991wr, Leibovich:1997az} is given by
\begin{align}
\sigma(w)= \frac{1}{\sqrt{(w+1)(w-1)}}
\langle
  {qq (1^{-}, 0)}|{qq (0^{+}, 0)}
\rangle\,.
\end{align}

The normalization of the IW functions for transitions into ground-state hadrons is fixed at zero recoil, $w = 1$. 
Accordingly, they can be expanded as
\begin{align}
H(w) = 1 + \lambda_{H} (w-1) + \cdots
\,,
\end{align}
where $H = \xi,\,\zeta$, and $\lambda_H$ denotes the slope parameter.
In contrast, in the heavy-quark limit, the matrix elements for transitions from ground-state hadrons to excited states vanish at zero recoil due to the orthogonality of the light degrees of freedom, and their squares scale as $(w-1)$ in the vicinity of $w = 1$~\cite{Isgur:1990jf, Isgur:1991wr, Falk:1991nq, Leibovich:1997tu, Leibovich:1997em}.
Correspondingly, the IW functions are parametrized as
\begin{align}
H(w)
&=
\eta_{H} [1 + \lambda_{H} (w-1) + \cdots]
\,,
\end{align}
where $H = \tau_{1/2},\,\tau_{3/2},\, \sigma$, and $\eta_H$ encodes the overall normalization.

By multiplying kinematic factors and performing the phase-space integration over three decay angles, Eq.~\eqref{eq:amplitude} leads to a relation among the differential decay rates,
\begin{align}
 \frac{\kappa_{\Lambda_c}}{\zeta(w)^2} &= 
 \frac{2}{w+1}\frac{\kappa_{D}+\kappa_{D^*}}{\xi(w)^2} \nonumber \\
 &= \frac{1}{(w+1)(w-1)}\frac{\kappa_{\Lambda_c^*(1/2^-)}+\kappa_{\Lambda_c^*(3/2^-)}}{\sigma(w)^2} 
 \nonumber \\
 &= \frac{1}{2(w-1)}\frac{\kappa_{D_0^*}+\kappa_{D_1^*}}{\tau_{1/2}(w)^2}
 = \frac{1}{(w+1)^2(w-1)}\frac{\kappa_{D_1}+\kappa_{D_2^*}}{\tau_{3/2}(w)^2}  \,,
 \label{eq:kappa_SR1}
\end{align}
where $\kappa_{H_c}\equiv d\Gamma(H_b \to H_c\tau\ov\nu )/dw$.
This relation is exact in the heavy-quark limit for arbitrary NP contributions described by Eq.~\eqref{eq:Hamiltonian}.
The coefficients $\gamma_i^{\,\mathrm{SV}}$ are obtained by integrating over the phase space in the SV limit.

We begin by briefly reviewing the case of decays into ground-state charm hadrons, $B\to D^{1/2^-}\tau\ov\nu$ and $\Lambda_b\to\Lambda_c\tau\ov\nu$, discussed in Ref.~\cite{Endo:2025lvy}. 
Integrating the first line of Eq.~\eqref{eq:kappa_SR1} over the phase space in the SV limit, in which $w_{\rm max}\to 1$ and the LO IW functions satisfy $\xi(w),\,\zeta(w)\to 1$, one finds $\Gamma_{\Lambda_c} = \Gamma_{D} + \Gamma_{D^*}$. 
In the same limit, one also has $\Gamma_{D^*}^{\,\mathrm{SM}} = 3\, \Gamma_{D}^{\,\mathrm{SM}}$, so that the coefficients reduce to the simple rational values,
\begin{align}
 \gamma_{2}^{\,\mathrm{SV}} = \frac{1}{4} \,, ~~~
 \gamma_{3}^{\,\mathrm{SV}} = \frac{3}{4} \,.
 \label{eq:SR_g2g_coeff}
\end{align}
This reproduces Eq.~\eqref{eq:3mode_2508} with $\delta = 0$.
For physical hadrons, however, these limits are not exact, and the deviation therefore becomes nonzero.
The sum rule is then given by 
\begin{align}
 \delta^{\mathrm{SV}}[\Lambda_c,D^{1/2^-}] = 
 \frac{R_{\Lambda_c}}{R_{\Lambda_c}^{\,\mathrm{SM}}} 
 - 
 \left[
 \gamma_{2}^{\,\mathrm{SV}}
 \frac{R_{D}}{R_{D}^{\,\mathrm{SM}}} 
 +
 \gamma_{3}^{\,\mathrm{SV}}
 \frac{R_{D^*}}{R_{D^*}^{\,\mathrm{SM}}} 
 \right]
 \label{eq:SR_g2g}
 \,,
\end{align}
where $\gamma_i^{\,\mathrm{SV}}$ are evaluated in the SV limit, yielding $\gamma_{2}^{\,\mathrm{SV}} = 1/4$ and $\gamma_{3}^{\,\mathrm{SV}} = 3/4$, as in Ref.~\cite{Endo:2025lvy}.

We apply this procedure to derive sum rules for decays into excited charm hadrons.
In contrast to transitions between ground-state hadrons, the matrix elements for transitions from ground-state hadrons into excited states vanish at zero recoil, $w = 1$.
This feature is reflected in the $1/(w-1)$ factors appearing in the second and third lines of Eq.~\eqref{eq:kappa_SR1}. 
Hence, by multiplying both sides of these expressions by $(w-1)$ and integrating over $w$, we find that the leading terms in the SV limit satisfy the relation (cf.~Ref.~\cite{Endo:2025lvy}),
\begin{align}
 \frac{1}{2\,\eta_{\,\sigma}^2} \left(\Gamma_{\Lambda_c^*(1/2^-)}+\Gamma_{\Lambda_c^*(3/2^-)}\right)
 = \frac{1}{2\,\eta_{\,\tau_{1/2}}^2} \left(\Gamma_{D_0^*}+\Gamma_{D_1^*}\right)
 = \frac{1}{4\,\eta_{\,\tau_{3/2}}^2} \left(\Gamma_{D_1}+\Gamma_{D_2^*}\right) 
 \,.
 \label{eq:GammaSV}
\end{align}
By normalizing to the corresponding SM predictions $\Gamma_{H_c}^{\mathrm{SM}}$, this relation can be recast as
\begin{align}
 \frac{\Gamma_{\Lambda_c^*(1/2^-)}+\Gamma_{\Lambda_c^*(3/2^-)}}
 {\Gamma_{\Lambda_c^*(1/2^-)}^{\,\mathrm{SM}}+\Gamma_{\Lambda_c^*(3/2^-)}^{\,\mathrm{SM}}} 
 =
 \frac{\Gamma_{D_0^*}+\Gamma_{D_1^*}}{\Gamma_{D_0^*}^{\,\mathrm{SM}}+\Gamma_{D_1^*}^{\,\mathrm{SM}}}
 = 
 \frac{\Gamma_{D_1}+\Gamma_{D_2^*}}{\Gamma_{D_1}^{\,\mathrm{SM}}+\Gamma_{D_2^*}^{\,\mathrm{SM}}}
 \,.
\end{align}
This implies the following relation among the total decay rates,
\begin{align}
 \alpha_{\Lambda_c^*}^{\mathrm{SV}}
 \frac{\Gamma_{\Lambda_c^*(1/2^-)}}{\Gamma_{\Lambda_c^*(1/2^-)}^{\,\mathrm{SM}}} +
 \beta_{\Lambda_c^*}^{\,\mathrm{SV}}
 \frac{\Gamma_{\Lambda_c^*(3/2^-)}}{\Gamma_{\Lambda_c^*(3/2^-)}^{\,\mathrm{SM}}}
 =
 \alpha_{D^{1/2^+}}^{\mathrm{SV}}
 \frac{\Gamma_{D_0^*}}{\Gamma_{D_0^*}^{\,\mathrm{SM}}} + 
 \beta_{D^{1/2^+}}^{\,\mathrm{SV}}
 \frac{\Gamma_{D_1^*}}{\Gamma_{D_1^*}^{\,\mathrm{SM}}}
 = 
 \alpha_{D^{3/2^+}}^{\mathrm{SV}}
 \frac{\Gamma_{D_1}}{\Gamma_{D_1}^{\,\mathrm{SM}}} + 
 \beta_{D^{3/2^+}}^{\,\mathrm{SV}}
 \frac{\Gamma_{D_2^*}}{\Gamma_{D_2^*}^{\,\mathrm{SM}}}
 \,,
 \label{eq:Gamma_excited}
\end{align}
where the coefficients are defined as
\begin{align}
 \label{eq:alphabeta_1}
 & \alpha_{\Lambda_c^*}^{\mathrm{SV}}
 = \frac{\Gamma_{\Lambda_c^*(1/2^-)}^{\,\mathrm{SM}}}
 {\Gamma_{\Lambda_c^*(1/2^-)}^{\,\mathrm{SM}}+\Gamma_{\Lambda_c^*(3/2^-)}^{\,\mathrm{SM}}}
 \,, ~~~
 \beta_{\Lambda_c^*}^{\,\mathrm{SV}}
 = \frac{\Gamma_{\Lambda_c^*(3/2^-)}^{\,\mathrm{SM}}}
 {\Gamma_{\Lambda_c^*(1/2^-)}^{\,\mathrm{SM}}+\Gamma_{\Lambda_c^*(3/2^-)}^{\,\mathrm{SM}}}
 \,, \\
 \label{eq:alphabeta_2}
 & \alpha_{D^{1/2^+}}^{\mathrm{SV}}
 = \frac{\Gamma_{D_0^*}^{\,\mathrm{SM}}}
 {\Gamma_{D_0^*}^{\,\mathrm{SM}}+\Gamma_{D_1^*}^{\,\mathrm{SM}}}
 \,, ~~~
 \beta_{D^{1/2^+}}^{\,\mathrm{SV}}
 = \frac{\Gamma_{D_1^*}^{\,\mathrm{SM}}}
 {\Gamma_{D_0^*}^{\,\mathrm{SM}}+\Gamma_{D_1^*}^{\,\mathrm{SM}}}
 \,, \\
 \label{eq:alphabeta_3}
 & \alpha_{D^{3/2^+}}^{\,\mathrm{SV}}
 = \frac{\Gamma_{D_1}^{\,\mathrm{SM}}}
 {\Gamma_{D_1}^{\,\mathrm{SM}}+\Gamma_{D_2^*}^{\,\mathrm{SM}}}
 \,, ~~~
 \beta_{D^{3/2^+}}^{\,\mathrm{SV}}
 = \frac{\Gamma_{D_2^*}^{\,\mathrm{SM}}}
 {\Gamma_{D_1}^{\mathrm{SM}}+\Gamma_{D_2^*}^{\,\mathrm{SM}}}
 \,.
\end{align}
By construction, $\alpha_{H_c}^{\mathrm{SV}}+\beta_{H_c}^{\,\mathrm{SV}}=1$ for $H_c = \Lambda_c^*, D^{1/2^+}, D^{3/2^+}$. 
Equation \eqref{eq:Gamma_excited} is trivially satisfied in the SM limit $\Gamma_{H_c} \to \Gamma_{H_c}^{\,\mathrm{SM}}$.

Finally, normalizing further to the total decay rates of the light-lepton modes $\Gamma(H_b\to H_c \ell\ov\nu)$, we obtain the sum rule relating decays into $P$-wave excited hadrons,
\begin{align}
 \alpha_{\Lambda_c^*}^{\mathrm{SV}}
 \frac{R_{\Lambda_c^*(1/2^-)}}{R_{\Lambda_c^*(1/2^-)}^{\,\mathrm{SM}}} +
 \beta_{\Lambda_c^*}^{\,\mathrm{SV}}
 \frac{R_{\Lambda_c^*(3/2^-)}}{R_{\Lambda_c^*(3/2^-)}^{\,\mathrm{SM}}}
 = 
 \alpha_{D^{1/2^+}}^{\mathrm{SV}}
 \frac{R_{D_0^*}}{R_{D_0^*}^{\,\mathrm{SM}}} +
 \beta_{D^{1/2^+}}^{\,\mathrm{SV}}
 \frac{R_{D_1^*}}{R_{D_1^*}^{\,\mathrm{SM}}} 
 = 
 \alpha_{D^{3/2^+}}^{\mathrm{SV}}
 \frac{R_{D_1}}{R_{D_1}^{\,\mathrm{SM}}} +
 \beta_{D^{3/2^+}}^{\,\mathrm{SV}}
 \frac{R_{D_2^*}}{R_{D_2^*}^{\,\mathrm{SM}}}
 \,.
 \label{eq:SR_excited}
\end{align}
This relation holds in the SV limit and remains valid even in the presence of NP contributions described by Eq.~\eqref{eq:Hamiltonian}. 

For realistic hadrons, however, the limit is not exact, and deviations arise from effects of hadron masses as well as corrections to the transition form factors (see Ref.~\cite{Endo:2025fke}).
These effects are parametrized by the deviation appearing in the sum rule of Eq.~\eqref{eq:4SR_general}. 
For instance, the sum rule among $\Lambda_b\to \Lambda_c^*\tau\ov\nu$ and $B\to D^{1/2^+}\tau\ov\nu$ takes the form, 
\begin{align}
 \delta^{\mathrm{SV}}[\Lambda_c^*,D^{1/2^+}] 
 &= 
 \left[
 \frac{R_{\Lambda_c^*(1/2^-)}}{R_{\Lambda_c^*(1/2^-)}^{\,\mathrm{SM}}} 
 +
 \gamma_{1}^{\,\mathrm{SV}}
 \frac{R_{\Lambda_c^*(3/2^-)}}{R_{\Lambda_c^*(3/2^-)}^{\,\mathrm{SM}}} 
 \right]
 - 
 \left[
 \gamma_{2}^{\,\mathrm{SV}}
 \frac{R_{D_0^*}}{R_{D_0^*}^{\,\mathrm{SM}}} 
 +
 \gamma_{3}^{\,\mathrm{SV}}
 \frac{R_{D_1^*}}{R_{D_1^*}^{\,\mathrm{SM}}}
 \right]
 \,, 
 \label{eq:SR_SV_Lamcst_D12p}
\end{align}
where the coefficients $\gamma_i^{\,\mathrm{SV}}$ are taken to be
\begin{align}
\gamma_1^{\,\mathrm{SV}} 
= 
\frac{\beta_{\Lambda_c^*}^{\,\mathrm{SV}}}
{\alpha_{\Lambda_c^*}^{\mathrm{SV}}}\,, 
\qquad
\gamma_2^{\,\mathrm{SV}} 
= 
\frac{\alpha_{D^{1/2^+}}^{\mathrm{SV}}}
{\alpha_{\Lambda_c^*}^{\mathrm{SV}}}\,, 
\qquad 
\gamma_3^{\,\mathrm{SV}} 
= 
\frac{\beta_{D^{1/2^+}}^{\,\mathrm{SV}}}
{\alpha_{\Lambda_c^*}^{\mathrm{SV}}}\,.
\end{align}
In the SV limit, this relation reproduces Eq.~\eqref{eq:SR_excited} with $\delta^{\mathrm{SV}}[\Lambda_c^*,D^{1/2^+}] = 0$.
The sum rules for the other combinations, 
$\Lambda_b\to \Lambda_c^*\tau\ov\nu$ and $B\to D^{3/2^+}\tau\ov\nu$, 
as well as 
$B\to D^{1/2^+}\tau\ov\nu$ and $B\to D^{3/2^+}\tau\ov\nu$, 
can be constructed analogously. 

Once the SV limit is no longer exact, there is no unique prescription for determining the corrections to the coefficients $\gamma_{i}^{\rm SV}$. 
For the sum rule relating transitions between ground-state hadrons in Eq.~\eqref{eq:SR_g2g}, the coefficients are still evaluated in the SV limit, since they reduce to simple rational values given in Eq.~\eqref{eq:SR_g2g_coeff}.
For transitions into excited hadrons, however, the coefficients in Eqs.~\eqref{eq:alphabeta_1}--\eqref{eq:alphabeta_3} do not reduce to such simple values.
We therefore define them according to Eqs.~\eqref{eq:alphabeta_1}--\eqref{eq:alphabeta_3} and evaluate them using the SM predictions for physical hadrons.
With this choice, Eq.~\eqref{eq:gamma_constraint_4} continues to hold even away from the SV limit.

In the above, we considered sum rules involving only decays into excited hadrons.
The SV-limit prescription cannot be used to construct sum rules relating decays into ground-state hadrons to those into excited states.
The reason is that the matrix elements of the light degrees of freedom exhibit different behavior near $w=1$.
More specifically, the matrix elements for transitions into ground-state hadrons remain finite at zero recoil, whereas those for transitions into excited states vanish.
As a result, even after multiplying the first line of Eq.~\eqref{eq:kappa_SR1} by $(w-1)$ and integrating over $w$, the result cannot be written in terms of the total decay rates for $B \to D^{1/2^-}\tau\ov\nu$ and $\Lambda_b\to\Lambda_c\tau\ov\nu$. 
Sum rules for such cases therefore require a different prescription and will be discussed in Appendices~\ref{app:KIT_coefficients} and \ref{app:SR_g2g_vs_g2e}.

\subsection{KIT prescription}
\label{sec:SR_KIT}

The KIT prescription~\cite{Blanke:2018yud, Blanke:2019qrx, Fedele:2022iib} does not rely on the relation in Eq.~\eqref{eq:kappa_SR1} or on the SV limit. 
Instead, the coefficients $\gamma_{i}^{\,\mathrm{KIT}}$ in Eqs.~\eqref{eq:4SR_general} and \eqref{eq:3SR_general} are fixed such that selected NP contributions, in addition to the SM one, are removed from the deviation $\delta^{\mathrm{KIT}}[H_c,H_c']$. 
Eliminating a pair of operator contributions $(kl)$ amounts to imposing $\delta^{\mathrm{KIT}}_{kl}[H_c,H_c'] = 0$, in addition to $\delta^{\mathrm{KIT}}_{V_L V_L}[H_c,H_c'] = 0$, the latter being equivalent to Eqs.~\eqref{eq:gamma_constraint_4} and~\eqref{eq:gamma_constraint_3}. 
The number of operator pairs $(kl)$ that can be eliminated simultaneously matches the number of remaining free coefficients $\gamma_{i}^{\,\mathrm{KIT}}$, namely two for Eq.~\eqref{eq:4SR_general} and one for Eq.~\eqref{eq:3SR_general}. 
In this work, we adopt $(kl) = (V_L S_R)$ and $(V_L S_L)$ for Eq.~\eqref{eq:4SR_general}, and $(kl) = (V_L S_L)$ for Eq.~\eqref{eq:3SR_general} as a reference choice.\footnote{
We have checked that alternative choices of $(kl)$ lead to qualitatively similar results and do not affect the conclusions of this work.
}
That is, we impose
\begin{align}
 \delta^{\mathrm{KIT}}_{V_L S_R}[H_c,H_c'] 
 = \delta^{\mathrm{KIT}}_{V_L S_L}[H_c,H_c'] 
 = 0
\label{eq:KIT_reference_choice_4SR}
\end{align}
for Eq.~\eqref{eq:4SR_general}, and 
\begin{align}
 \delta^{\mathrm{KIT}}_{V_L S_L}[H_c,H_c'] 
 = 0
\label{eq:KIT_reference_choice_3SR}
\end{align}
for Eq.~\eqref{eq:3SR_general}.
Solving Eqs.~\eqref{eq:gamma_constraint_4} and \eqref{eq:KIT_reference_choice_4SR}, or Eqs.~\eqref{eq:gamma_constraint_3} and \eqref{eq:KIT_reference_choice_3SR}, determines the coefficients $\gamma_{i}^{\,\mathrm{KIT}}$.
Their explicit expressions in both cases are given in Appendix~\ref{app:KIT_coefficients}.

\subsection{Cancellation measure}

A sum rule is useful when the corresponding deviation $\delta^X[H_c,H_c']$ is small enough. 
However, since $\delta^X[H_c,H_c']$ involves the Wilson coefficients as in Eq.~\eqref{eq:deltaX}, a small deviation can result simply from small Wilson coefficients, regardless of the structure of the sum rule. 
Moreover, even at the level of individual $\delta_{ij}^X[H_c,H_c']$, a small value can indicate either that the individual ratios $R^{\,ij}_{H_c}/R_{H_c}^{\,\mathrm{SM}}$ themselves are small, or that efficient cancellations occur among them. 
Only the latter probes the non-trivial cancellation built into the sum rule.
Consequently, $\delta_{ij}^{X}[H_c,H_c']$ do not in general provide a reliable measure of the degree of cancellation. 

We therefore introduce a measure that quantifies the relative size of the cancellation,
\begin{align}
  \epsilon_{ij}^X[H_c,H_c'] = 
  \frac{\big|\,\delta_{ij}^X[H_c,H_c']\,\big|}
  { \max
  \left[\,\,
  \Bigg| \dfrac{R_{H_c(1)}^{ij}}{R_{H_c(1)}^{\,\mathrm{SM}}} \Bigg|,\,
  \Bigg| \gamma_{1}^{X} \dfrac{R_{H_c(2)}^{\,ij}}{R_{H_c(2)}^{\,\mathrm{SM}}} \Bigg|,\,
  \Bigg| \gamma_{2}^{X} \dfrac{R_{H_c'(1)}^{\,ij}}{R_{H_c'(1)}^{\,\mathrm{SM}}} \Bigg|,\,
  \Bigg| \gamma_{3}^{X} \dfrac{R_{H_c'(2)}^{\,ij}}{R_{H_c'(2)}^{\,\mathrm{SM}}} \Bigg|
  \,\,\right] }
  \,.
  \label{eq:epsilon}
\end{align}
This definition applies when both $H_c$ and $H_c'$ belong to heavy-quark doublets. 
If $H_c$ is a heavy-quark singlet, the measure is given by
\begin{align}
  \epsilon_{ij}^X[H_c,H_c'] = 
  \frac{\big|\,\delta_{ij}^X[H_c,H_c']\,\big|}
  { \max
  \left[\,\,
  \Bigg| \dfrac{R_{H_c}^{ij}}{R_{H_c}^{\,\mathrm{SM}}} \Bigg|,\,
  \Bigg| \gamma_{2}^{X} \dfrac{R_{H_c'(1)}^{\,ij}}{R_{H_c'(1)}^{\,\mathrm{SM}}} \Bigg|,\,
  \Bigg| \gamma_{3}^{X} \dfrac{R_{H_c'(2)}^{\,ij}}{R_{H_c'(2)}^{\,\mathrm{SM}}} \Bigg|
  \,\,\right] }
  \,.
  \label{eq:epsilon_singlet}
\end{align}
Efficient cancellation then corresponds to $\epsilon_{ij}^X[H_c,H_c']\ll 1$, while $\epsilon_{ij}^X[H_c,H_c'] = \mathcal{O}(1)$ indicates poor cancellation.
The sum rule is therefore useful when $\epsilon_{ij}^X[H_c,H_c']\ll 1$ and $\delta^X[H_c,H_c']$ is smaller than the experimental uncertainties.

\section{Numerical result}
\label{sec:Numerical}

In this section, we study the deviation $\delta_{ij}^X$ and the cancellation measure $\epsilon_{ij}^X$ of the sum rules in order to assess their quantitative predictive power. 
In particular, one of the ratios $R_{H_c}$ entering the relation can be predicted using the experimental inputs for the remaining ratios together with the SM expectations. 
For instance, $R_{H_c(1)}$ can be obtained from Eqs.~\eqref{eq:4SR_general} as\footnote{
  Predictions for $R_{H_c(2)}$, $R_{H_c'(1)}$, and $R_{H_c'(2)}$ can be constructed analogously from Eq.~\eqref{eq:pred}.
} 
\begin{align}
 R_{H_c(1)} = 
 R_{H_c(1)}^{\rm SM}
 \left\{ 
 \delta^X[H_c,H_c'] -
 \gamma_{1}^X
 \frac{R_{H_c(2)}}{R_{H_c(2)}^{\rm SM}} + 
 \gamma_{2}^X
 \frac{R_{H_c'(1)}}{R_{H_c'(1)}^{\rm SM}} +
 \gamma_{3}^X
 \frac{R_{H_c'(2)}}{R_{H_c'(2)}^{\rm SM}}
 \right\} 
 \,.
 \label{eq:pred}
\end{align}
The coefficients $\gamma_{i}^X$ are independent of the NP operators, while the NP dependence enters explicitly through the deviation $\delta^X$.
Therefore, if $\delta^X$ is sufficiently small compared to the experimental uncertainties, the prediction becomes effectively insensitive to NP contributions.
More precisely, in order to assess whether the sum rules provide a non-trivial consistency test of the experimental data and the SM predictions, we compare $\delta^X$ with the relative uncertainty in the prediction for $R_{H_c(1)}$, normalized to its SM value.
The current status and projected experimental precisions are summarized in Table~\ref{tab:prospect}.

The decay rates are evaluated using the physical hadron masses.
In addition, we employ form factors based on the heavy-quark effective theory (HQET).
Exploiting the heavy-quark symmetry in $b \to c$ transitions~\cite{Isgur:1989vq, Isgur:1990yhj, Isgur:1990pm, Georgi:1990cx}, form factors associated with different effective operators can be expressed in terms of a common set of IW functions.\footnote{
In contrast, in model-independent approaches such as the BGL parametrization~\cite{Boyd:1995cf}, form factors are parametrized independently, rather than being reduced to a common set of IW functions by imposing heavy quark symmetry.
}
For instance, tensor form factors can be related to vector form factors at leading order~\cite{Neubert:1993mb, Falk:1992wt, Falk:1992ws, Bernlochner:2017jxt, Leibovich:1997az}.
Higher-order corrections in the heavy-quark expansion are also incorporated in the analysis. 

The form factor parameters are determined using experimental data, theoretical constraints, and lattice QCD inputs.\footnote{
For decays into ground-state hadrons, we follow Refs.~\cite{Bernlochner:2017jka, Bordone:2019vic, Iguro:2020cpg} for $B\to D^{(*)}\tau\ov\nu$ and Refs.~\cite{Bernlochner:2018kxh} for $\Lambda_b\to\Lambda_c\tau\ov\nu$.
}
For $B\to D^{**}\tau\ov\nu$, we use Ref.~\cite{Bernlochner:2017jxt}, which includes corrections up to $\mathcal{O}(\Lambda_{\rm QCD}/m_Q)$, where $\Lambda_{\rm QCD}$ denotes the QCD scale.
The form factor parameters are constrained by the differential distributions of $B\to D_0^* \ell\ov\nu$ and $B\to D_2^* \ell\ov\nu$, as well as the branching fraction of $B\to D^{3/2^+}\pi$, under the assumption of QCD factorization.
For $\Lambda_b\to\Lambda_c^*\tau\ov\nu$, we use Ref.~\cite{Papucci:2021pmj}, where the form factors are constrained using the relative branching ratios between excited and ground states, together with lattice QCD inputs.
At present, however, these constraints do not allow a fully data-driven determination of higher-order contributions~\cite{Bernlochner:2017jxt,Papucci:2021pmj}.
More precise measurements of light-lepton modes by Belle II and LHCb are expected in the future~\cite{Belle-II:2018jsg,LHCb:2018roe}.
In the numerical results presented below, we use the central values of the form factor parameters.
Accordingly, the deviations and cancellation measures should be interpreted as central-value estimates.
We will return to this issue at the end of this section and discuss how the form factor uncertainties affect the interpretation of the numerical results. 

\begin{figure}[t]
\begin{center}
\includegraphics[width=0.25\linewidth]{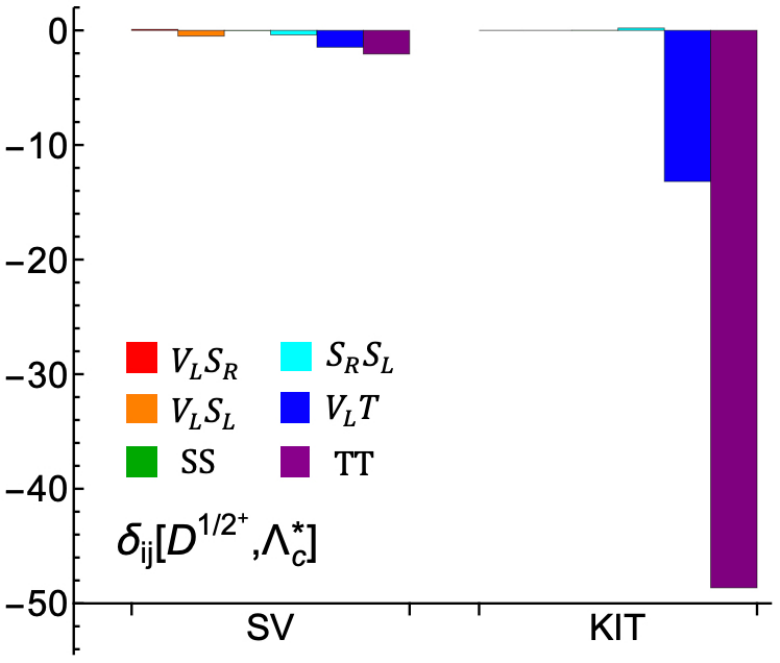}~~
\includegraphics[width=0.25\linewidth]{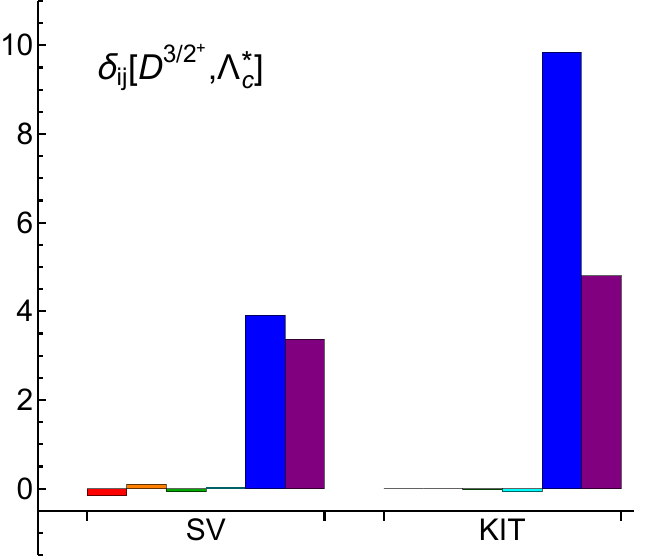}~~
\includegraphics[width=0.25\linewidth]{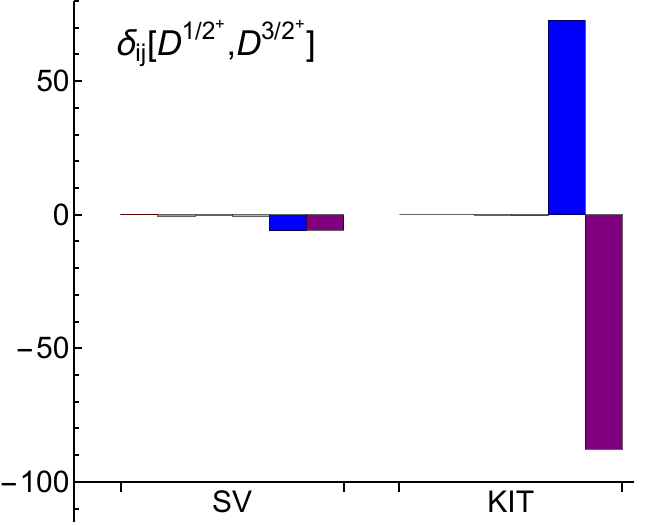}\\\vspace{0.25cm}
\includegraphics[width=0.25\linewidth]{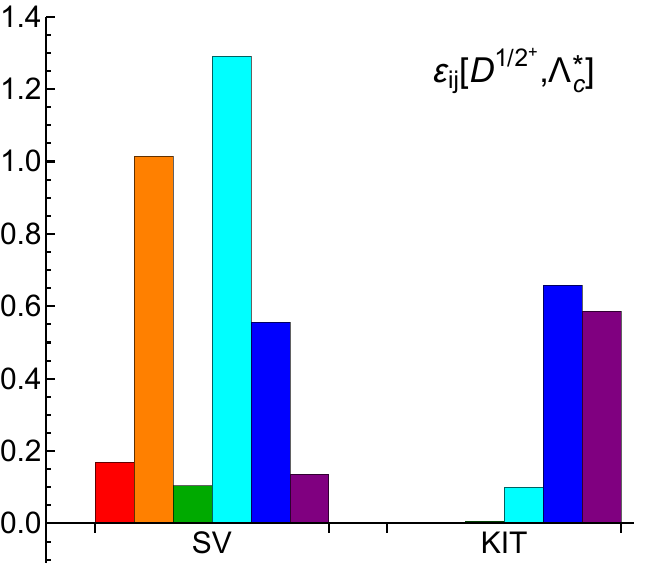}~~
\includegraphics[width=0.25\linewidth]{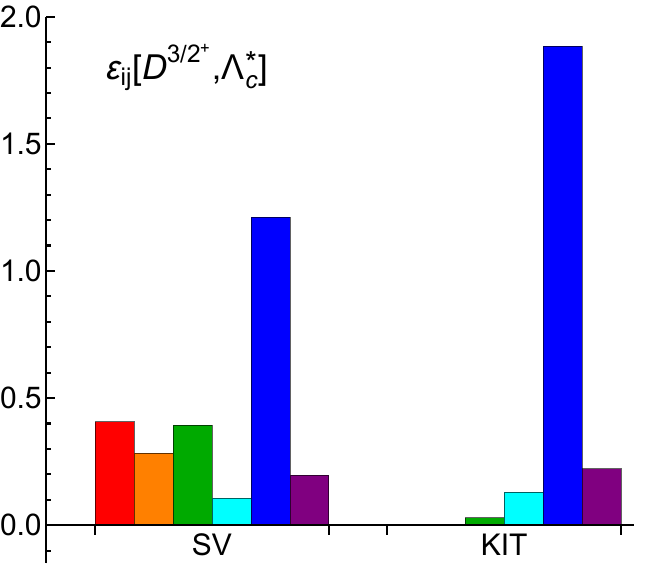}~~
\includegraphics[width=0.25\linewidth]{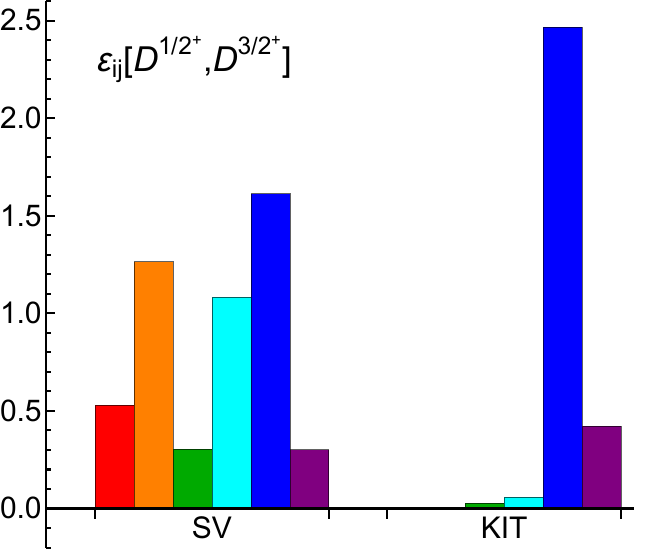}
\end{center}
\vspace{-.15cm}
\caption{
The deviations $\delta_{ij}^X$ (upper) and cancellation measures $\epsilon_{ij}^X$ (lower) for the sum rules relating pairs of decays into excited-state doublets.
The left, middle, and right columns correspond to the combinations of decay channels $(D^{1/2^+},\,\Lambda_c^*)$, $(D^{3/2^+},\,\Lambda_c^*)$, and $(D^{1/2^+},\,D^{3/2^+})$, respectively.
For each panel, the two groups labeled ``SV'' and ``KIT'' show the results obtained with the SV-limit prescription and the KIT prescription for the sum rule coefficients, respectively.
}
\label{fig:4mode_int_ee}
\end{figure}

In Fig.~\ref{fig:4mode_int_ee}, we present the numerical results for $\delta_{ij}^X$ and $\epsilon_{ij}^X$ in the sum rules relating decays into excited hadrons, $B \to D^{1/2^+}\tau\ov\nu$ (labeled $D^{1/2^+}$ in the plot and caption), $B \to D^{3/2^+}\tau\ov\nu$ (labeled $D^{3/2^+}$), and $\Lambda_b \to \Lambda_c^*\tau\ov\nu$ (labeled $\Lambda_c^*$).
From these three decay channels, one can construct three independent sum rules. 
As seen in the figure, the deviation $\delta_{ij}^X$ becomes sizable in all cases when the NP contributions involve tensor operators (blue and purple bars). 
However, the cancellation measure $\epsilon_{ij}^X$ can reach $\mathcal{O}(1)$ even in other scenarios, such as those involving scalar contributions. 
This indicates that violations of the SV limit can degrade the validity of the sum rules.

When the coefficients are determined by the SV-limit prescription (labeled ``SV''), certain NP contributions can significantly spoil the sum rules, particularly in the presence of scalar or tensor operators.
We identify two possible sources of such violations.
First, the SV limit itself may be significantly violated.
Since the form factors are suppressed near $w = 1$, the total decay rates become sensitive to their behavior away from zero recoil. 
Second, the form factors are not yet precisely determined. 
In particular, the $V_L S_{R,L}$ ($V_L T$) contributions shown in the plot arise from the interference between vector and scalar (tensor) operators, and therefore depend on multiple form factors.
These form factors may be subject to systematic uncertainties. 
This issue is especially relevant for transitions into excited charm hadrons, for which the available form factor information remains limited~\cite{Bernlochner:2017jxt, Papucci:2021pmj}.
The conclusions involving excited modes should therefore be revisited once more precise determinations of the form factors become available.

In contrast, in the KIT prescription (labeled ``KIT''), we impose the elimination of the $V_L V_L$, $V_L S_R$ (red), and $V_L S_L$ (orange) contributions. 
The $SS$ (green) and $S_R S_L$ (cyan) contributions are then also reduced simultaneously.
Nevertheless, the sum rules involving tensor contributions (blue and purple) remain sensitive to these violations and can be significantly degraded.
This conclusion should therefore be reassessed in the future, once more precise form factor inputs are available.

\begin{table}[t]
\begin{center}
 \scalebox{0.95}{
  \begin{tabular}{cccc} 
 Scenario & Parameter & Value &  Pull  \\ \hline
$S_L$ & $C_{S_L}$ & $-0.57\pm0.86\,i$ & 4.3\\
$S_R$ & $C_{S_R}$ & 0.18 & 3.9\\
$T$ & $C_{T}$ & $0.02 \pm 0.13\,i$ & 3.8\\\hdashline
${\rm{R}}_2$ & $C_{S_L}=8.4\,C_T$ & $-0.09 \pm 0.56\,i$ & 4.4\\
${\rm{S}}_1$ & $C_{S_L}=-8.9\,C_T$ & $0.18$ & 4.1\\
${\rm{U}}_1$ & $C_{V_L}$,\,$\phi$ & $0.075,\,\pm 0.466\pi$ & 4.4\\ \hline
\end{tabular}
}
  \caption{Fit results for the Wilson coefficients in single-operator ($S_L$, $S_R$, $T$) and single leptoquark (${\rm R}_2$, ${\rm S}_1$, ${\rm U}_1$) scenarios.
  The first column indicates the scenario, with the relevant coefficients listed in the second column.
  For the ${\rm U}_1$ leptoquark, we consider $U(2)$-flavored scenario, which satisfies the relation $C_{S_R} = -3.7e^{i\phi}C_{V_L}$.
  The best-fit values of the coefficients at the $\mu_b$ scale are presented in the third column, and the fit quality, the pull value is shown in the fourth column.
  See Ref.~\cite{Iguro:2024hyk} for further details.
 }
  \label{Tab:scenario}
\end{center}   
\vspace{-.15cm}
\end{table}

In the above analysis, the deviation has been decomposed into NP contributions and expressed in terms of the corresponding Wilson coefficients.
These coefficients are constrained by experimental data, for example in attempts to account for the current discrepancies between the measured values and the SM predictions.
In the following, we focus on the $R_{D^{(*)}}$ anomalies~\cite{HFLAV:2024ctg}.
In particular, we consider three ``single-operator'' scenarios and three ``single-leptoquark'' scenarios (see Ref.~\cite{Iguro:2024hyk} for details).
In the former case, NP effects are described within the low-energy effective field-theory framework, in which a single operator is assumed to be nonzero while all others vanish.
In the latter case, the discrepancies are addressed by the exchange of a single leptoquark (LQ).
In both cases, the corresponding Wilson coefficients are summarized in Table~\ref{Tab:scenario}.
They are obtained from a global fit to the experimental data on $R_{D^{(*)}}$ and the $D^*$ longitudinal polarization $F_L^{D^*}$, as described in Ref.~\cite{Iguro:2024hyk}.

\begin{figure}[t]
\begin{center}
\includegraphics[width=0.26\linewidth]{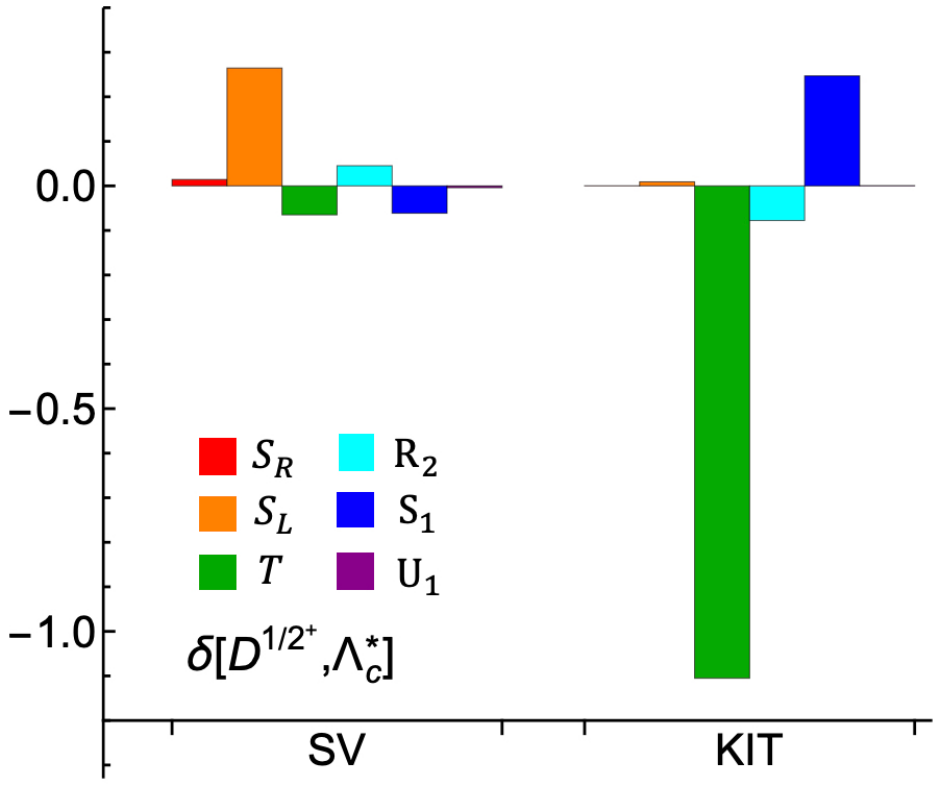}~~
\includegraphics[width=0.26\linewidth]{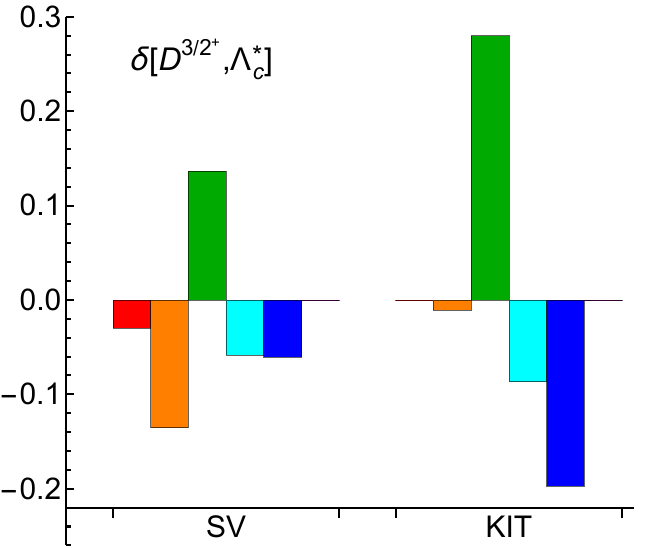}~~
\includegraphics[width=0.26\linewidth]{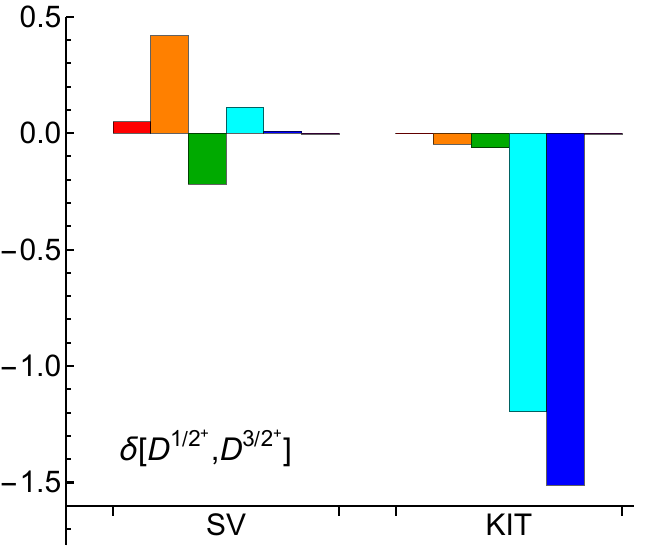}
\end{center}
\vspace{-.15cm}
\caption{The sum rule violation $\delta^X$ in the NP scenarios for the sum rules relating pairs of decays into excited-state hadrons.
The left, middle, and right panels correspond to the combinations of decay channels $(D^{1/2^+},\,\Lambda_c^*)$, $(D^{3/2^+},\,\Lambda_c^*)$, and $(D^{1/2^+},\,D^{3/2^+})$, respectively.
In each panel, the two groups labeled ``SV'' and ``KIT'' show the results obtained with the SV-limit prescription and the KIT prescription for the sum rule coefficients, respectively.
}
\label{fig:NP_4mode_e2e}
\end{figure}

In Fig.~\ref{fig:NP_4mode_e2e}, we present the deviation $\delta^X$ for the sum rules relating decays into excited hadrons. 
The coefficients entering the sum rules are determined either from the SV limit or within the KIT prescription.
We find that $|\delta^X|$ can exceed $0.1$ when the $S_L$ or $T$ operator accounts for the $R_{D^{(*)}}$ anomalies in the single-operator scenario. 
In contrast to Fig.~\ref{fig:4mode_int_ee}, the deviation becomes sizable also in the $S_L$ scenario, because the corresponding Wilson coefficient is favored to be large, as shown in Table~\ref{Tab:scenario}.
In the single LQ scenario, sizable $S_L$ or $T$ contributions can arise from $R_2$ or $S_1$.
Comparing with the projected sensitivities in Table~\ref{tab:prospect}, such large deviations may compromise the robustness of the sum rules, particularly for the relation involving $B \to D^{3/2^+}\tau\ov\nu$ and $\Lambda_b \to \Lambda_c^*\tau\ov\nu$.
Given the current limited precision of the form factors for transitions into excited hadrons, however, this conclusion should be regarded as tentative and may need to be revisited in the future.

So far, we have used the central values of the form factor parameters and have not included the associated uncertainties.
At present, higher-order terms in the form factor parametrizations are not yet sufficiently constrained for transitions into excited charm hadrons~\cite{Bernlochner:2017jxt,Papucci:2021pmj}.
Accordingly, uncertainties in $R^{ij}_{H_c}/R^{\rm SM}_{H_c}$ can become sizable, and may even be of $\mathcal{O}(1)$ or not reliably assessable for some scalar and tensor contributions involving $D^{1/2^+}$ and $\Lambda_c^*$.
The deviations estimated above should therefore be regarded as indicative measures of the possible size of sum rule violations.
Nevertheless, more precise measurements of decay distributions in light-lepton modes are expected in the future, which will improve the determination of the form factors~\cite{Belle-II:2018jsg,LHCb:2018roe}.
Once such high-precision data become available, it will be worthwhile to revisit these sum rules.
 
\section{Conclusion}
\label{sec:Conclusion}

In this paper, we have studied semileptonic sum rules for $b \to c \tau \ov\nu$ transitions involving orbitally excited charm hadrons.
Motivated by the conventional sum rule among the ground-state ratios $R_{\Lambda_c}$, $R_D$, and $R_{D^*}$, we examined whether analogous relations can be constructed for decays into the excited meson doublets $D^{1/2^+}$ and $D^{3/2^+}$ and the excited baryon doublet $\Lambda_c^*$, and whether such relations remain useful once realistic hadronic effects are taken into account.

Starting from the amplitude-level relation implied by heavy-quark symmetry, we constructed sum rules relating decays into excited hadrons in the SV limit. 
Since the coefficients of these relations are no longer uniquely fixed once the SV limit is violated, we also introduced the KIT prescription, in which the coefficients are chosen to suppress selected operator contributions.
In the appendix, we also explored sum rules relating decays into ground-state hadrons and those into excited hadrons.
Since a construction in the SV limit is not available at the level of total decay rates, we adopted the KIT prescription and studied its phenomenological consequences.

Once excited hadrons are included, the resulting sum rules remain qualitatively well motivated, though our numerical analysis, based on the central values of the form-factor parameters, suggests that their quantitative robustness is reduced.
In representative NP scenarios, the deviation can exceed the level of $\mathcal{O}(0.1)$, which is comparable to the precision target of future experimental tests. 
Given that future measurements of $R_{H_c}$ may achieve precisions at the level of $\lesssim 10\%$, further improvements in theoretical predictions are required for these relations to serve as reliable consistency checks.

The primary limitation of the present analysis arises from the current knowledge of the form factors for transitions into excited hadrons. 
In particular, tensor form factors are not yet well constrained. 
At present, higher-order contributions in the form-factor parametrizations remain insufficiently determined, which limits the quantitative reliability of the sum rules. 
As a result, the deviations obtained in this work should be regarded as indicative measures of the possible size of sum rule violations.

More precise measurements of light-lepton decay distributions, together with improved theoretical determinations of the form factors, will be essential for establishing more robust relations involving excited charm hadrons. 
Once such inputs become available, the sum rules studied here should provide useful consistency tests of semileptonic $b \to c\tau\ov\nu$ transitions, including possible NP effects.

\section*{Acknowledgements}
%
We appreciate Abhijit Mathad, Ryoutaro Watanabe and Hiroyasu Yonaha for stimulating discussion.
This work is supported by JSPS KAKENHI Grant Numbers 22K21347 [M.E. and S.I.], 24K07025 [S.M.], 24K22879 [S.I.], 24K23939 [S.I.], 25K17385 [S.I.] and Toyoaki scholarship foundation [S.I.].
We also appreciate KEK-KMI joint appointment program [M.E. and S.I.]. 
\appendix

\section{Fitting formula}
\label{app:fitting_formula}
The fitting formulae for mesonic and baryonic decays are obtained using the form factor inputs described in Sec.~\ref{sec:Numerical} as
\begin{align}
 \label{eq:RD}
 \frac{R_D}{R_{D}^\textrm{SM}} =
 & ~|1+C_{V_L}|^2  + 1.00|C_{S_R}+C_{S_L}|^2 + 0.81|C_{T}|^2  \nonumber \\[-0.5em]
 & + 1.49\textrm{Re}[(1+C_{V_L})(C_{S_R}^*+C_{S_L}^*)]  + 1.07\textrm{Re}[(1+C_{V_L})C_{T}^*] \,, 
 \\[0.4em]
 \label{eq:RDst}
 \frac{ R_{D^*}}{R_{D^*}^\textrm{SM}} =
 & ~|1+C_{V_L}|^2  + 0.04|C_{S_R}-C_{S_L}|^2 + 16.0|C_{T}|^2 \nonumber \\[-0.5em]
 &  + 0.12\textrm{Re}[(1+C_{V_L})(C_{S_R}^*-C_{S_L}^*)] -5.13\textrm{Re}[(1+C_{V_L})C_{T}^*]  \,, 
 \\[0.4em]
 \label{eq:RDs0}
 \frac{R_{D^*_0}}{R_{D^*_0}^\textrm{SM}} =
 & ~|1+C_{V_L}|^2  + 0.15|C_{S_R}-C_{S_L}|^2 + 2.70|C_{T}|^2  \nonumber \\[-0.5em]
 & + 0.49\textrm{Re}[(1+C_{V_L})(C_{S_R}^*-C_{S_L}^*)]  -2.41\textrm{Re}[(1+C_{V_L})C_{T}^*] \,, 
 \\[0.4em]
 \label{eq:RDs1}
 \frac{R_{D^*_1}}{R_{D^*_1}^\textrm{SM}} =
 & ~|1+C_{V_L}|^2  + 0.25|C_{S_R}+C_{S_L}|^2 + 31.0|C_{T}|^2  \nonumber \\[-0.5em]
 & + 0.52\textrm{Re}[(1+C_{V_L})(C_{S_R}^*+C_{S_L}^*)]  -3.85\textrm{Re}[(1+C_{V_L})C_{T}^*] \,, 
 \\[0.4em]
 \label{eq:RD1}
 \frac{R_{D_1}}{R_{D_1}^\textrm{SM}} =
 & ~|1+C_{V_L}|^2  + 0.18|C_{S_R}+C_{S_L}|^2 + 17.3|C_{T}|^2  \nonumber \\[-0.5em]
 & + 0.38\textrm{Re}[(1+C_{V_L})(C_{S_R}^*+C_{S_L}^*)]  +3.23\textrm{Re}[(1+C_{V_L})C_{T}^*] \,, 
 \\[0.4em]
 \label{eq:RDs2}
 \frac{R_{D^*_2}}{R_{D^*_2}^\textrm{SM}} =
 & ~|1+C_{V_L}|^2  + 0.04|C_{S_R}-C_{S_L}|^2 + 11.7|C_{T}|^2  \nonumber \\[-0.5em]
 & + 0.14\textrm{Re}[(1+C_{V_L})(C_{S_R}^*-C_{S_L}^*)]  -5.79\textrm{Re}[(1+C_{V_L})C_{T}^*] \,,
\end{align} 
\begin{align}
\frac{R_{\Lambda_c}}{R_{\Lambda_c}^{\rm SM}}& =  |1+C_{V_L}|^2 
+ 0.47 \,\textrm{Re}[ (1 +C_{V_L}) C_{S_R}^*]
+ 0.30 \,\textrm{Re}[ (1 +C_{V_L}) C_{S_L}^* ]
+ 0.46 \, \textrm{Re}[C_{S_L} C_{S_R}^* ]
\nonumber \\[0.35em]
&\quad  + 0.29 \, (|C_{S_L}|^2 + |C_{S_R}|^2) -3.36 \,\textrm{Re}[ (1 +C_{V_L})  C_T^*] + 12.2 \, |C_T|^2\,,\\[1em]
\frac{R_{\Lambda_c^*(1/2^-)}}{R_{\Lambda_c^*(1/2^-)}^{\rm SM}}& =  |1+C_{V_L}|^2 
+ 0.43 \,\textrm{Re}[ (1 +C_{V_L}) C_{S_R}^*]
+ 0.04 \,\textrm{Re}[ (1 +C_{V_L}) C_{S_L}^* ]
+ 0.10 \, \textrm{Re}[C_{S_L} C_{S_R}^* ]
\nonumber \\[0.3em]
&\quad  + 0.18 \, (|C_{S_L}|^2 + |C_{S_R}|^2) -0.27 \,\textrm{Re}[ (1 +C_{V_L})  C_T^*] + 13.6 \, |C_T|^2\,,\\[1em]
\frac{R_{\Lambda_c^*(3/2^-)}}{R_{\Lambda_c^*(3/2^-)}^{\rm SM}}& =  |1+C_{V_L}|^2 
+ 0.45 \,\textrm{Re}[ (1 +C_{V_L}) C_{S_R}^*]
+ 0.28 \,\textrm{Re}[ (1 +C_{V_L}) C_{S_L}^* ]
+ 0.32 \, \textrm{Re}[C_{S_L} C_{S_R}^* ]
\nonumber \\[0.3em]
&\quad  + 0.21 \, (|C_{S_L}|^2 + |C_{S_R}|^2) -3.19 \,\textrm{Re}[ (1 +C_{V_L})  C_T^*] + 13.2 \, |C_T|^2\,.
\end{align}
The SM predictions are given by 
\begin{align}
R_{\Lambda_c}^{\rm SM}\sim\,\,&0.32\,,\,\,\,
R_{D}^{\rm SM}\sim0.29\,,\,\,\,
R_{D^*}^{\rm SM}\sim0.25\,,\,\,\,
R_{D_0^*}^{\rm SM}\sim0.08\,,\,\,\,
R_{D_1^*}^{\rm SM}\sim0.05\,,\nonumber\\
R_{D_1}^{\rm SM}\sim\,\,&0.10\,,\,\,\,
R_{D_2^*}^{\rm SM}\sim0.07\,,\,\,\,
R_{\Lambda_c^*(1/2^-)}^{\rm SM}\sim0.17\,,\,\,\,
R_{\Lambda_c^*(3/2^-)}^{\rm SM}\sim0.11\,.
\end{align}

\section{Explicit expressions for the KIT coefficients}
\label{app:KIT_coefficients}

In this appendix, we provide the explicit expressions for the coefficients $\gamma_{i}^{\,\mathrm{KIT}}$ appearing in the KIT prescription, adopting the reference choice specified in Eqs.~\eqref{eq:KIT_reference_choice_4SR} and~\eqref{eq:KIT_reference_choice_3SR}. 
For other operator choices, the corresponding expressions are obtained by replacing $V_L S_R$ and $V_L S_L$ with the eliminated operator pairs. 
We introduce
\begin{align}
 a^{ij}_{H_c} \equiv \frac{R^{\,ij}_{H_c}}{R_{H_c}^{\,\mathrm{SM}}} \,,
\end{align}
for simplicity. 

For the sum rule of Eq.~\eqref{eq:4SR_general}, in which both $H_c$ and $H_c'$ are heavy-quark doublets, the three coefficients $\gamma_{1,2,3}^{\,\mathrm{KIT}}$ are determined by imposing $\delta^{\,\mathrm{KIT}}_{V_L V_L}[H_c,H_c'] = 0$, $\delta^{\,\mathrm{KIT}}_{V_L S_R}[H_c,H_c'] = 0$ and $\delta^{\,\mathrm{KIT}}_{V_L S_L}[H_c,H_c'] = 0$, yielding
\begin{align}
 \gamma_{1}^{\,\mathrm{KIT}} 
 &=
 -\frac{a^{V_L S_R}_{H_c'(1)} a^{V_L S_L}_{H_c(1)} - a^{V_L S_R}_{H_c'(2)} a^{V_L S_L}_{H_c(1)} - a^{V_L S_R}_{H_c(1)} a^{V_L S_L}_{H_c'(1)} + a^{V_L S_R}_{H_c'(2)} a^{V_L S_L}_{H_c'(1)} + a^{V_L S_R}_{H_c(1)} a^{V_L S_L}_{H_c'(2)} - a^{V_L S_R}_{H_c'(1)} a^{V_L S_L}_{H_c'(2)}}{\mathcal{D}}
 \,, \nonumber \\[+0.4em]
 \gamma_{2}^{\,\mathrm{KIT}} 
 &=
 -\frac{a^{V_L S_R}_{H_c(2)} a^{V_L S_L}_{H_c(1)} - a^{V_L S_R}_{H_c'(2)} a^{V_L S_L}_{H_c(1)} - a^{V_L S_R}_{H_c(1)} a^{V_L S_L}_{H_c(2)} + a^{V_L S_R}_{H_c'(2)} a^{V_L S_L}_{H_c(2)} + a^{V_L S_R}_{H_c(1)} a^{V_L S_L}_{H_c'(2)} - a^{V_L S_R}_{H_c(2)} a^{V_L S_L}_{H_c'(2)}}{\mathcal{D}}
 \,, \nonumber \\[+0.4em]
 \gamma_{3}^{\,\mathrm{KIT}} 
 &=
 -\frac{-a^{V_L S_R}_{H_c(2)} a^{V_L S_L}_{H_c(1)} + a^{V_L S_R}_{H_c'(1)} a^{V_L S_L}_{H_c(1)} + a^{V_L S_R}_{H_c(1)} a^{V_L S_L}_{H_c(2)} - a^{V_L S_R}_{H_c'(1)} a^{V_L S_L}_{H_c(2)} - a^{V_L S_R}_{H_c(1)} a^{V_L S_L}_{H_c'(1)} + a^{V_L S_R}_{H_c(2)} a^{V_L S_L}_{H_c'(1)}}{\mathcal{D}}
 \,, 
 \label{eq:gamma_KIT_4_app}
\end{align}
with the common denominator 
\begin{align}
 \mathcal{D} = 
 a^{V_L S_R}_{H_c'(1)} a^{V_L S_L}_{H_c(2)} 
 - a^{V_L S_R}_{H_c'(2)} a^{V_L S_L}_{H_c(2)} 
 - a^{V_L S_R}_{H_c(2)} a^{V_L S_L}_{H_c'(1)} 
 + a^{V_L S_R}_{H_c'(2)} a^{V_L S_L}_{H_c'(1)} 
 + a^{V_L S_R}_{H_c(2)} a^{V_L S_L}_{H_c'(2)} 
 - a^{V_L S_R}_{H_c'(1)} a^{V_L S_L}_{H_c'(2)}
 \,.
 \label{eq:gamma_KIT_4_app_denominator}
\end{align}
The relation $1 + \gamma_{1}^{\,\mathrm{KIT}} = \gamma_{2}^{\,\mathrm{KIT}} + \gamma_{3}^{\,\mathrm{KIT}}$ can be verified, consistent with Eq.~\eqref{eq:gamma_constraint_4}. 

For the sum rule of Eq.~\eqref{eq:3SR_general}, in which $H_c$ is a heavy-quark singlet and $H_c'$ a doublet, the conditions $\delta^{\,\mathrm{KIT}}_{V_L V_L}[H_c,H_c'] = 0$ and $\delta^{\,\mathrm{KIT}}_{V_L S_L}[H_c,H_c'] = 0$ lead to 
\begin{align}
 \gamma_{2}^{\,\mathrm{KIT}} 
 &= 
 \frac{a^{V_L S_L}_{H_c} - a^{V_L S_L}_{H_c'(2)}}
 {a^{V_L S_L}_{H_c'(1)} - a^{V_L S_L}_{H_c'(2)}}
 \,,\qquad
 \gamma_{3}^{\,\mathrm{KIT}} 
 =
 \frac{a^{V_L S_L}_{H_c'(1)} - a^{V_L S_L}_{H_c}}
 {a^{V_L S_L}_{H_c'(1)} - a^{V_L S_L}_{H_c'(2)}}
 \,, 
 \label{eq:gamma_KIT_3_app}
\end{align}
where the relation $\gamma_{2}^{\,\mathrm{KIT}} + \gamma_{3}^{\,\mathrm{KIT}} = 1$ in Eq.~\eqref{eq:gamma_constraint_3} holds. 

The numerical values of the coefficients are summarized in Tables~\ref{tab:coefficients_g2e}, \ref{tab:coefficients_g2g_g2e}, and \ref{tab:coefficients_g2g_g2e_singlet}.
Table~\ref{tab:coefficients_g2e} also includes the corresponding values obtained in the SV-limit prescription.

\begin{table}[h]
   \centering
 \scalebox{0.89}{
\begin{tabular}{l|ccc}
      & $(D^{1/2^+},\,\Lambda_c^*)$ & $(D^{3/2^+},\,\Lambda_c^*)$ & $(D^{1/2^+},\,D^{3/2^+})$ \\
    \hline
    SV-limit & (0.49,\,0.68,\,0.81) & (0.31,\,0.60,\,0.71) & (0.49,\,1.14,\,0.35) \\
    KIT & (-2.04,\,5.24,\,-6.28) & (-0.28,\,-0.92,\,1.64) & (-6.72,\,-9.16,\,3.44) \\
    \hline
\end{tabular}
}
  \vspace{.15cm}
  \caption{Coefficients, ($\gamma_1^X,\,\gamma_2^X,\,\gamma_3^X$), for the sum rules relating pairs of decays into excited-state doublets.}
  \label{tab:coefficients_g2e}
\end{table}

\begin{table}[h]
   \centering
 \scalebox{0.89}{
\begin{tabular}{l|ccc}
      & $(D^{1/2^-},\,\Lambda_c^*)$ & $(D^{1/2^-},\,D^{1/2^+})$ & $(D^{1/2^-},\,D^{3/2^+})$ \\
    \hline
    KIT & (3.17,\,0.33,\,3.84) & (2.49,\,0.63,\,2.86) & (21.7,\,3.89,\,18.81) \\
    \hline
\end{tabular}
}
  \vspace{.15cm}
  \caption{Coefficients, ($\gamma_1^X,\,\gamma_2^X,\,\gamma_3^X$), for the sum rules relating $B \to D^{1/2^-}\tau\ov\nu$ to decays into excited hadrons.
  Here, $D^{1/2^-}$ belongs to a heavy-quark doublet.}
  \label{tab:coefficients_g2g_g2e}
\end{table}

\begin{table}[h]
  \centering
  \scalebox{0.95}{
\begin{tabular}{c|c|ccc}
      & $(\Lambda_c,\,D^{1/2^-})$ & $(\Lambda_c,\,\Lambda_c^*)$ & $(\Lambda_c,\,D^{1/2^+})$ & $(\Lambda_c,\,D^{3/2^+})$ \\
    \hline
    KIT & (0.27,\,0.73) & (-0.09,\,1.09) & (0.21,\,0.79) & (0.85,\,0.15) \\
    \hline
\end{tabular}
    }
  \vspace{.15cm}
  \caption{Coefficients, ($\gamma_1^X,\,\gamma_2^X$), for the sum rules relating $\Lambda_b \to \Lambda_c\tau\ov\nu$ to decays into excited hadrons, shown in the third to fifth columns.
  Here, $\Lambda_c$ belongs to a heavy-quark singlet.
  For comparison, the results for transitions between ground-state hadrons are shown in the second column.
  For the ground-state case, the SV-limit prescription gives $(\gamma_1^X,\gamma_2^X)=(0.25,0.75)$.
  \vspace{.3cm}
  }
  \label{tab:coefficients_g2g_g2e_singlet}
\end{table}

\section{Sum rule relating decays into ground-state hadrons and those into excited hadrons}
\label{app:SR_g2g_vs_g2e}

Although the sum rules derived in Sec.~\ref{sec:Sum_rule} involve only decays into excited hadrons, the experimental precision of such modes is still significantly worse than that for decays into ground-state hadrons, $B \to D^{1/2^-}\tau\ov\nu$ and $\Lambda_b\to\Lambda_c\tau\ov\nu$ (see Table~\ref{tab:prospect}).
In addition, as discussed in Sec.~\ref{sec:SR_SV}, the form factors for transitions into excited charm hadrons are currently less well constrained than those for ground-state transitions.
It is therefore useful to construct relations that connect decays into ground-state charm hadrons with decays into excited charm hadrons.

As noted in Sec.~\ref{sec:SR_SV}, the SV-limit prescription cannot be applied to such combinations.
In contrast, since the KIT prescription does not rely on the SV limit, we apply this method in the same way as in Sec.~\ref{sec:SR_KIT}.
For instance, the sum rule between $B\to D^{1/2^-}\tau\ov\nu$ and $B\to D^{1/2^+}\tau\ov\nu$ reads
\begin{align}
 \delta^{\,\mathrm{KIT}}[D^{1/2^-},D^{1/2^+}] 
 &= 
 \left[
 \frac{R_{D}}{R_{D}^{\,\mathrm{SM}}} 
 +
 \gamma_{1}^{\,\mathrm{KIT}}
 \frac{R_{D^*}}{R_{D^*}^{\,\mathrm{SM}}} 
 \right]
 - 
 \left[
 \gamma_{2}^{\,\mathrm{KIT}}
 \frac{R_{D_0^*}}{R_{D_0^*}^{\,\mathrm{SM}}} 
 +
 \gamma_{3}^{\,\mathrm{KIT}}
 \frac{R_{D_1^*}}{R_{D_1^*}^{\,\mathrm{SM}}}
 \right]
 \,, 
 \label{eq:SR_KIT_D12m_D12p}
\end{align}
while the sum rule between $\Lambda_b\to\Lambda_c\tau\ov\nu$ and $B\to D^{1/2^+}\tau\ov\nu$ is given by
\begin{align}
 \delta^{\,\mathrm{KIT}}[\Lambda_c,D^{1/2^+}] 
 &= 
 \frac{R_{\Lambda_c}}{R_{\Lambda_c}^{\,\mathrm{SM}}} 
 - 
 \left[
 \gamma_{2}^{\,\mathrm{KIT}}
 \frac{R_{D_0^*}}{R_{D_0^*}^{\,\mathrm{SM}}} 
 +
 \gamma_{3}^{\,\mathrm{KIT}}
 \frac{R_{D_1^*}}{R_{D_1^*}^{\,\mathrm{SM}}}
 \right]
 \,,
 \label{eq:SR_KIT_Lamc_D12p}
\end{align}
where the expressions of the coefficients $\gamma_i^{\,\mathrm{KIT}}$ are given in Appendix~\ref{app:KIT_coefficients}. 
The sum rules for the other combinations are constructed in the same way. 

\begin{figure}[t]
\begin{center}
\includegraphics[width=0.23\linewidth]{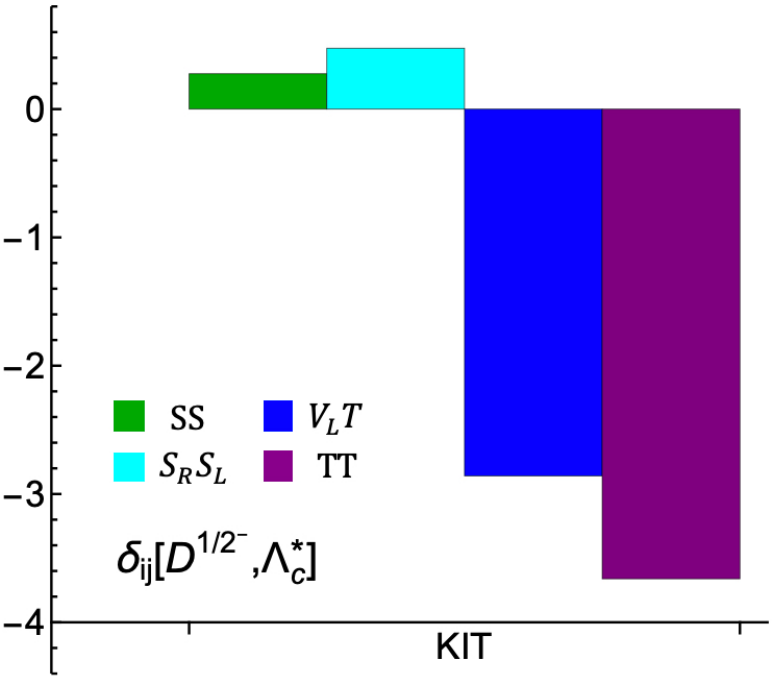}~
\includegraphics[width=0.23\linewidth]{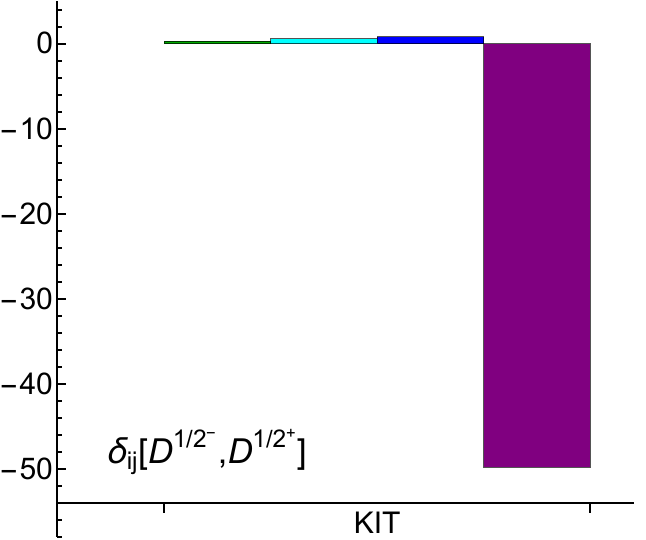}~
\includegraphics[width=0.23\linewidth]{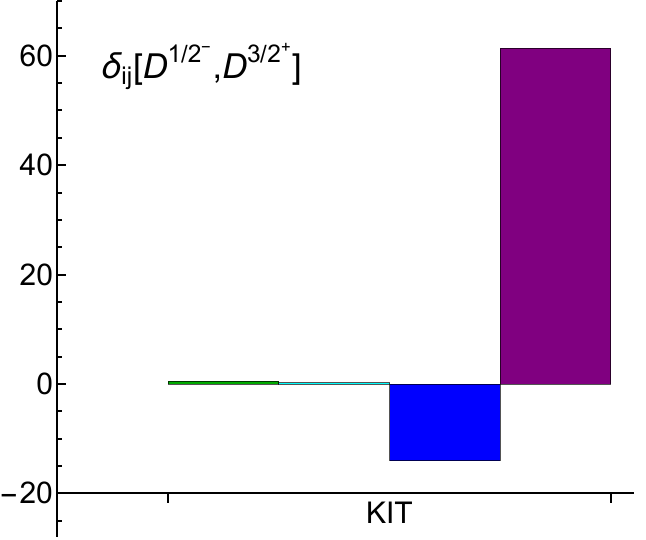}
\\\vspace{0.25cm}
\includegraphics[width=0.23\linewidth]{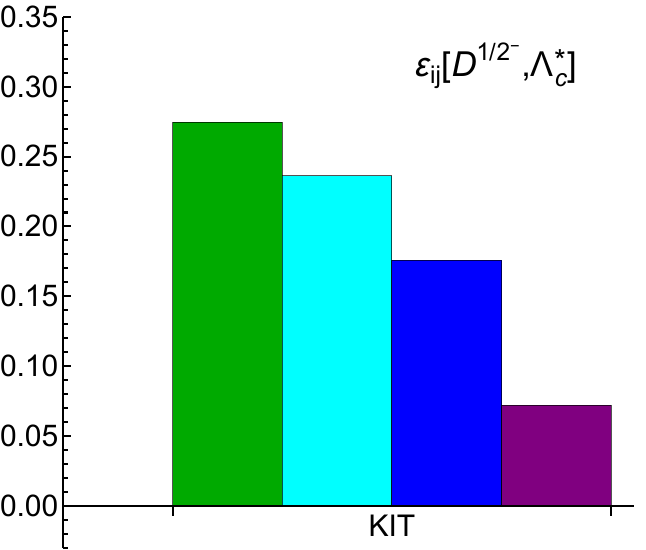}~
\includegraphics[width=0.23\linewidth]{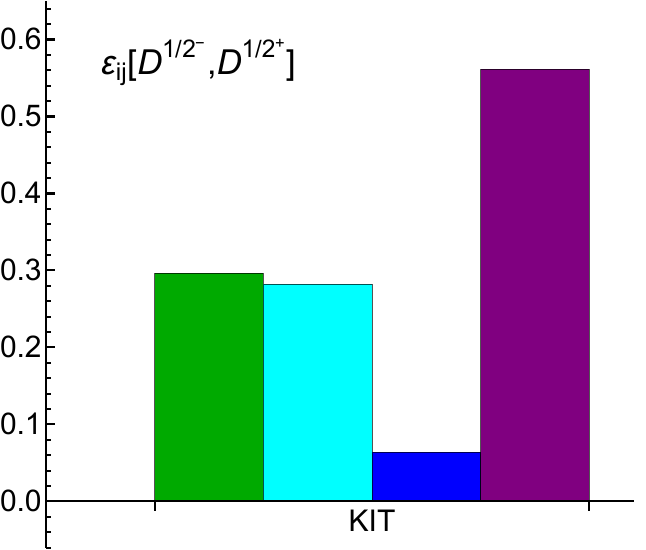}~
\includegraphics[width=0.23\linewidth]{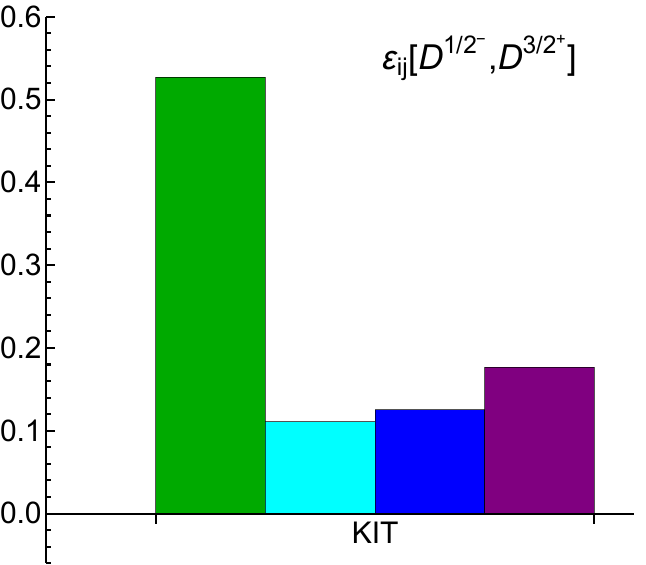}
\end{center}
\vspace{-.15cm}
\caption{The deviations $\delta_{ij}^X$ (upper) and cancellation measures $\epsilon_{ij}^X$ (lower) for the sum rules relating $B \to D^{1/2^-}\tau\ov\nu$ to decays into excited hadrons.
Here, $D^{1/2^-}$ belongs to a heavy-quark doublet.
The left, middle, and right columns correspond to the combinations of decay channels $(D^{1/2^-},\,\Lambda_c^*)$, $(D^{1/2^-},\,D^{1/2^+})$, and $(D^{1/2^-},\,D^{3/2^+})$, respectively.
The coefficients are determined within the KIT prescription, in which the $V_LV_L$, $V_LS_R$, and $V_LS_L$ contributions are eliminated.
}
\label{fig:4mode_int_ge}
\end{figure}

In Fig.~\ref{fig:4mode_int_ge}, we present the numerical results for the deviation $\delta_{ij}^X$ and the cancellation measure $\epsilon_{ij}^X$ in the sum rules relating $B \to D^{1/2^-}\tau\ov\nu$ to decays into excited hadrons.
Here, $D^{1/2^-}$ belongs to a heavy-quark doublet. 
In the KIT prescription, the coefficients of the sum rules are chosen such that the $V_L V_L$, $V_L S_R$, and $V_L S_L$ contributions are eliminated. 
In contrast to Fig.~\ref{fig:4mode_int_ee}, the $SS$ (green) and $S_R S_L$ (cyan) contributions are not generically suppressed.
Nevertheless, the cancellation measures remain below unity, indicating that the sum rules may remain effective even though they are not justified in the SV limit.

\begin{figure}[t]
\begin{center}
\includegraphics[width=0.23\linewidth]{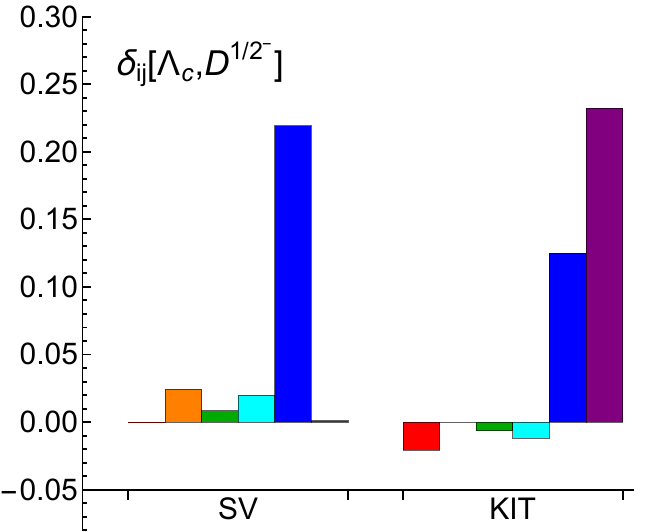}~
\includegraphics[width=0.23\linewidth]{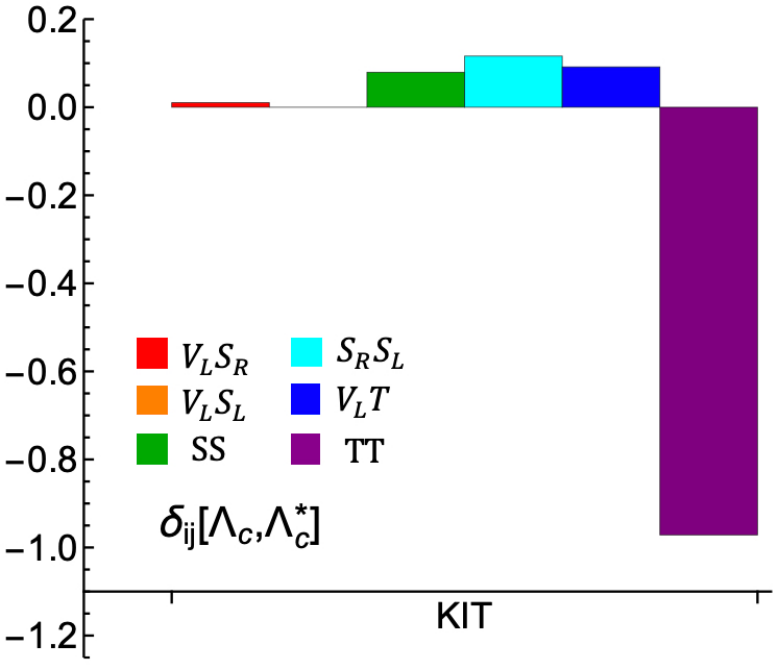}~
\includegraphics[width=0.23\linewidth]{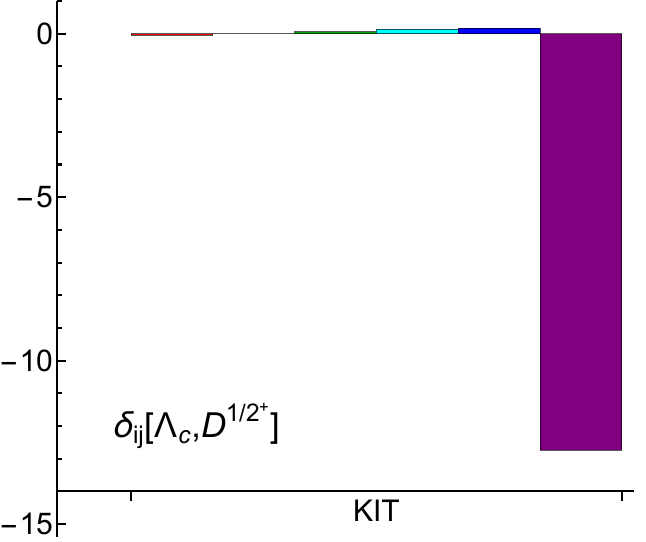}~
\includegraphics[width=0.23\linewidth]{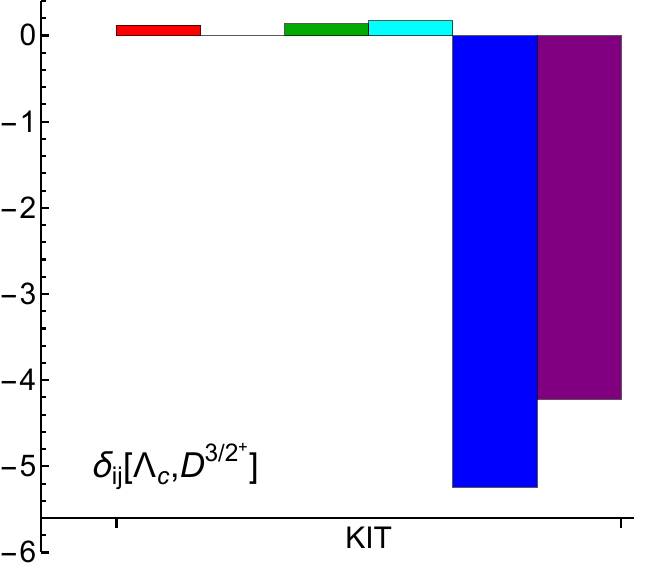}\\\vspace{0.25cm}
\includegraphics[width=0.23\linewidth]{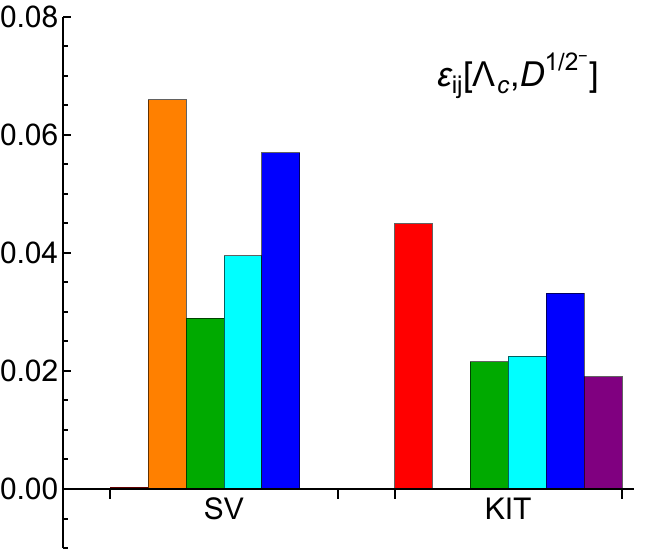}~
\includegraphics[width=0.23\linewidth]{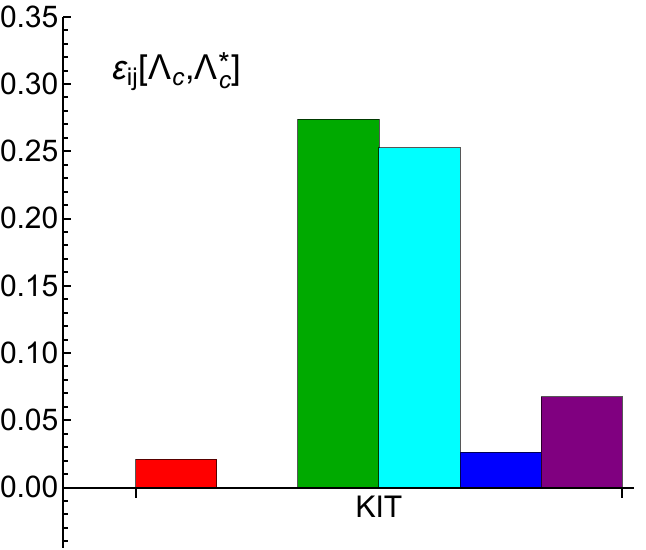}~
\includegraphics[width=0.23\linewidth]{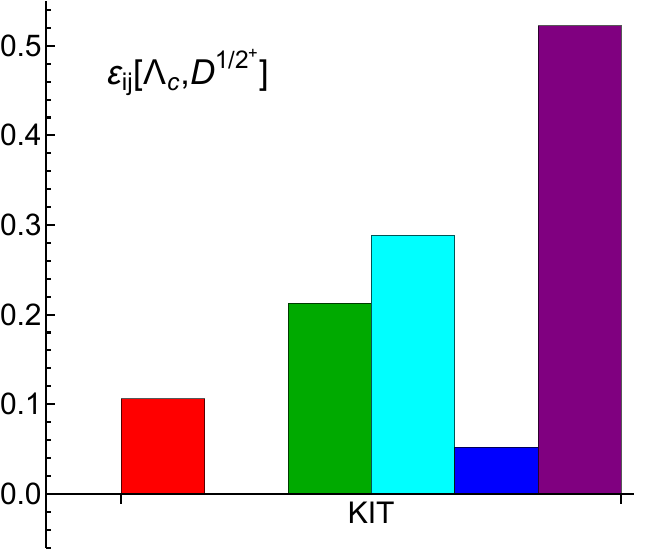}~
\includegraphics[width=0.23\linewidth]{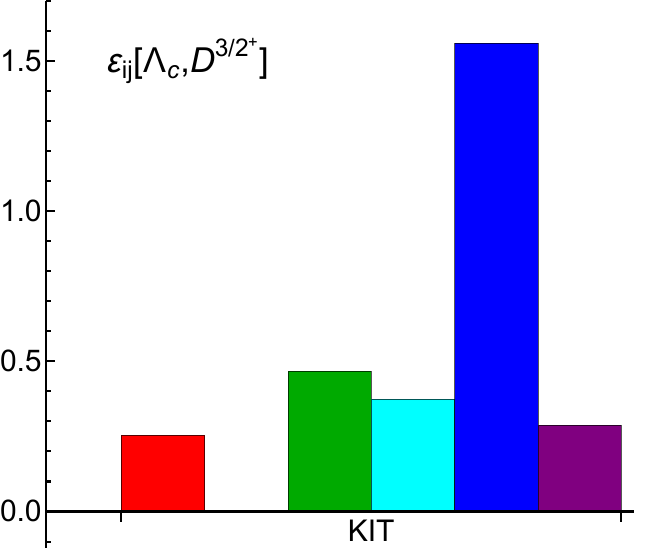}
\end{center}
\vspace{-.15cm}
\caption{The deviations $\delta_{ij}^X$ (upper) and cancellation measures $\epsilon_{ij}^X$ (lower).
The left-most column shows, for comparison, the sum rule relating the ground-state decays $\Lambda_b \to \Lambda_c \tau\ov\nu$ and $B \to D^{1/2^-} \tau\ov\nu$.
The second, third, and fourth columns correspond to the sum rules relating $\Lambda_b \to \Lambda_c \tau\ov\nu$ to $\Lambda_b \to \Lambda_c^* \tau\ov\nu$, $B \to D^{1/2^+} \tau\ov\nu$, and $B \to D^{3/2^+} \tau\ov\nu$, respectively.
For the left-most column, the results obtained with both the SV-limit prescription and the KIT prescription are shown.
For the other columns, the coefficients are determined within the KIT prescription, in which the $V_LV_L$ and $V_LS_L$ contributions are eliminated.
}
\label{fig:3mode_int_ge}
\end{figure}

In Fig.~\ref{fig:3mode_int_ge}, we show the results for the sum rules relating $\Lambda_b\to\Lambda_c\tau\ov\nu$ to decays into excited hadrons (except for the left-most panels, as discussed below).
Here, $\Lambda_c$ belongs to a heavy-quark singlet, and the coefficients of the sum rules are chosen within the KIT prescription such that the $V_L V_L$ and $V_L S_L$ contributions are eliminated. 
We find that the cancellation measures are well below unity, except for the $V_L T$ contribution (blue) for the sum rule between $\Lambda_b\to\Lambda_c\tau\ov\nu$ and $B \to D^{3/2^+}\tau\ov\nu$.
Therefore, at the level of the adopted central form factor inputs, the sum rules work reasonably well in most cases.

For comparison, the left-most panels show the results for transitions between ground-state hadrons, $\Lambda_b\to\Lambda_c\tau\ov\nu$ and $B \to D^{1/2^-}\tau\ov\nu$.
In this case, the coefficients entering the sum rule are determined either from the SV limit in Eq.~\eqref{eq:SR_g2g_coeff} or within the KIT prescription.
The deviations are significantly smaller than those involving decays into excited hadrons. 
In particular, unlike the latter case, the corresponding form factors are non-vanishing at zero recoil and have been determined with relatively high precision.  
This highlights the importance of improving the determination of form factors in order to draw robust conclusions.

\begin{figure}[t]
\begin{center}
\includegraphics[width=0.24\linewidth]{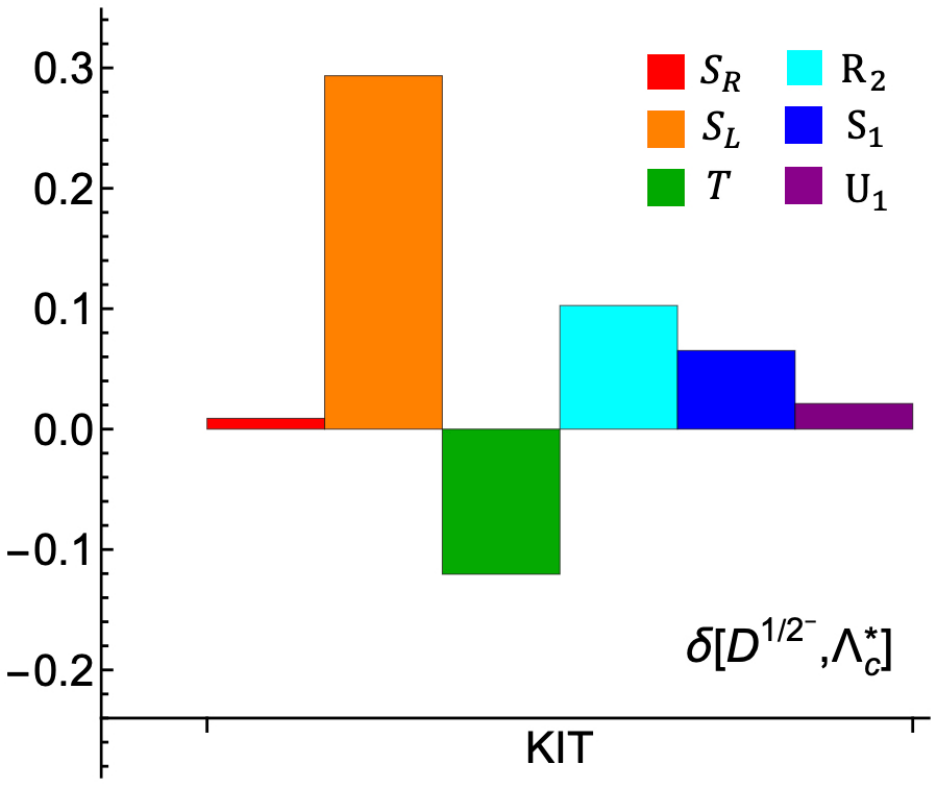}~~
\includegraphics[width=0.24\linewidth]{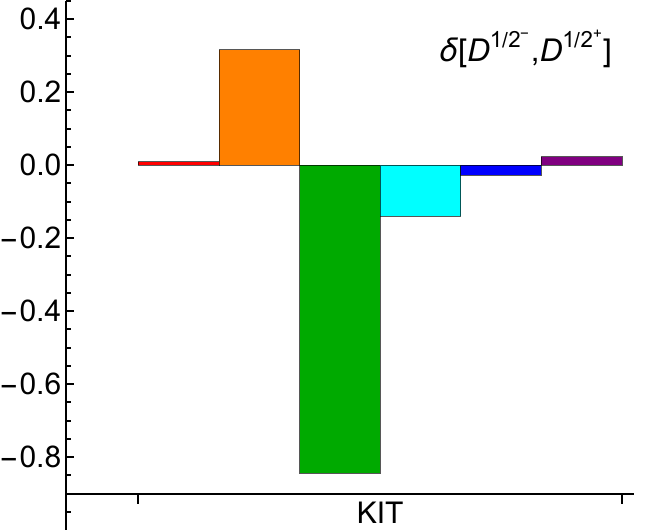}~~
\includegraphics[width=0.24\linewidth]{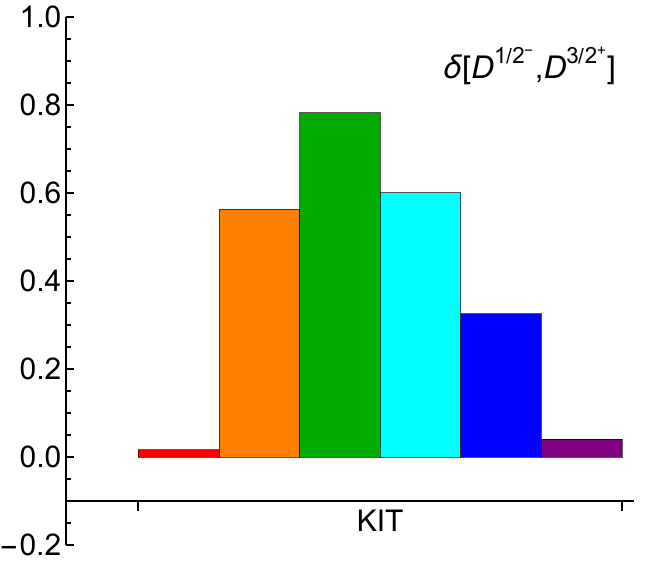}
\end{center}
\vspace{-.15cm}
\caption{The sum rule violation $\delta^X$ in the NP scenarios for the sum rules relating $B \to D^{1/2^-} \tau\ov\nu$ to decays into excited hadrons.
The left, middle, and right panels correspond to the combinations of decay channels $(D^{1/2^-},\,\Lambda_c^*)$, $(D^{1/2^-},\,D^{1/2^+})$, and $(D^{1/2^-},\,D^{3/2^+})$, respectively.
The coefficients are determined within the KIT prescription.
}
\label{fig:NP_4mode_g2e}
\end{figure}

\begin{figure}[t]
\begin{center}
\includegraphics[width=0.23\linewidth]{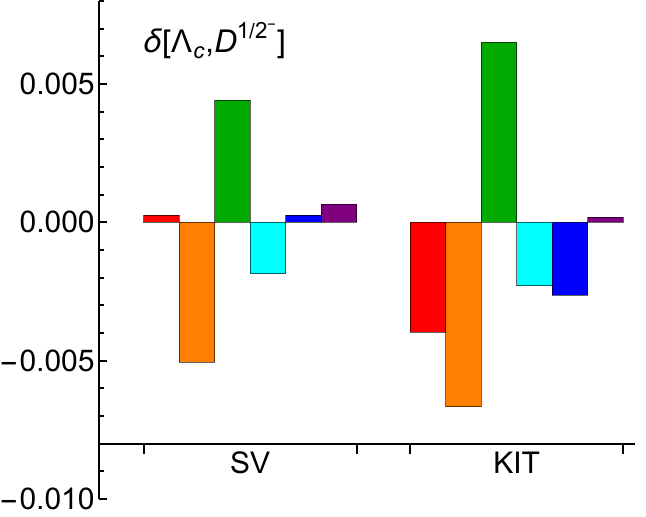}~
\includegraphics[width=0.23\linewidth]{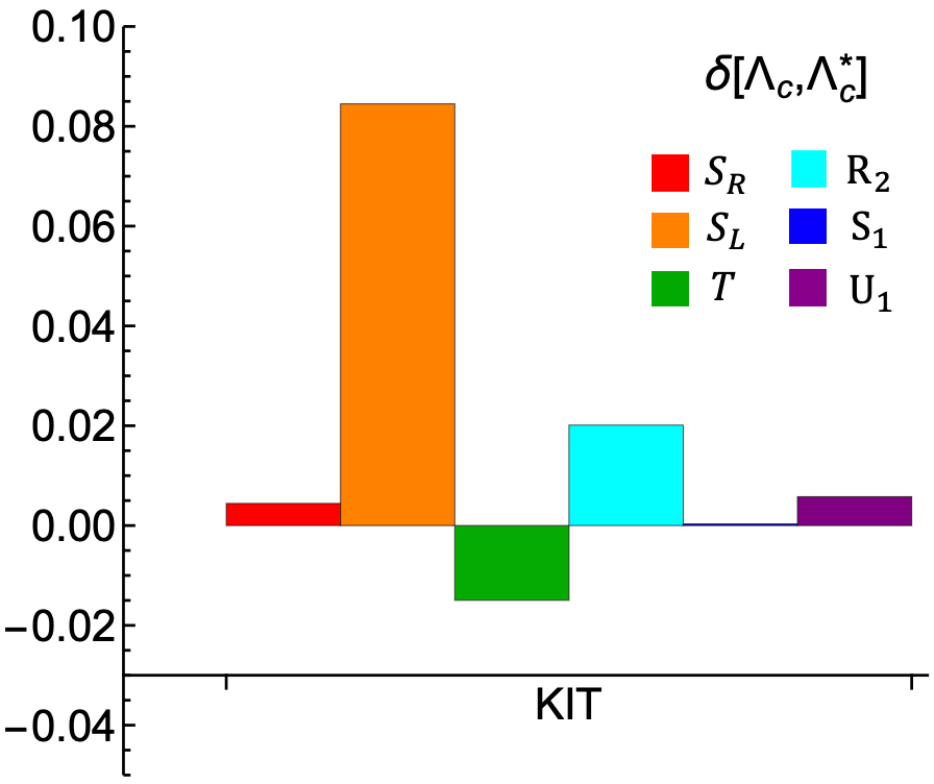}~
\includegraphics[width=0.23\linewidth]{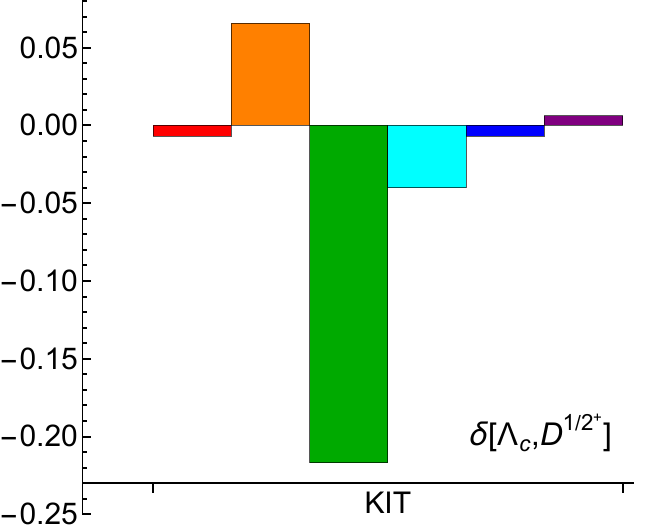}~
\includegraphics[width=0.23\linewidth]{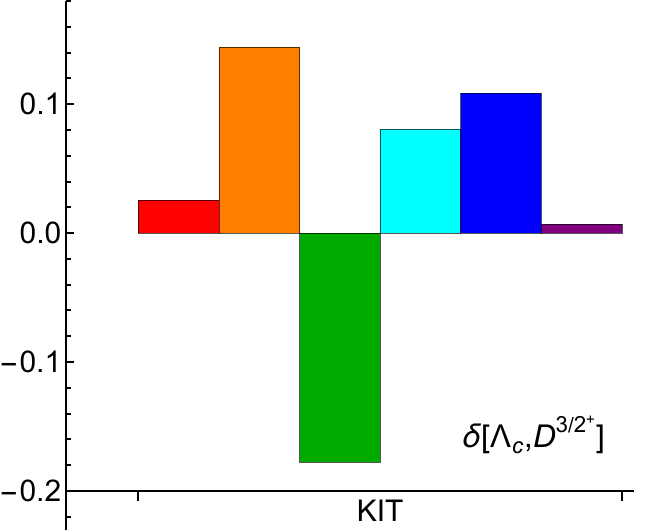}
\end{center}
\vspace{-.15cm}
\caption{The sum rule violation $\delta^X$ in the NP scenarios.
The left-most panel shows, for comparison, the sum rule relating the ground-state decays $\Lambda_b \to \Lambda_c \tau\ov\nu$ and $B \to D^{1/2^-} \tau\ov\nu$.
The second, third, and fourth panels correspond to the sum rules relating $\Lambda_b \to \Lambda_c \tau\ov\nu$ to $\Lambda_b \to \Lambda_c^* \tau\ov\nu$, $B \to D^{1/2^+} \tau\ov\nu$, and $B \to D^{3/2^+}\tau\ov\nu$, respectively.
For the left-most panel, both the SV-limit and KIT prescriptions are shown, whereas for the other panels the coefficients are determined within the KIT prescription.
}
\label{fig:NP_3mode_g2e}
\end{figure}

In Figs.~\ref{fig:NP_4mode_g2e} and \ref{fig:NP_3mode_g2e}, we evaluated $\delta^X$ within the NP scenarios shown in Table~\ref{Tab:scenario} for the sum rules relating $B \to D^{1/2^-}\tau\ov\nu$ and $\Lambda_b\to\Lambda_c\tau\ov\nu$ to decays into excited hadrons, respectively.
For comparison, the left-most panels in Fig.~\ref{fig:NP_3mode_g2e} show the results for transitions between ground-state hadrons, $\Lambda_b\to\Lambda_c\tau\ov\nu$ and $B \to D^{1/2^-}\tau\ov\nu$.
As in Fig.~\ref{fig:NP_4mode_e2e}, $|\delta^X|$ can exceed $0.1$ when the $S_L$ or $T$ operator accounts for the $R_{D^{(*)}}$ anomalies. 
Comparing with the projected sensitivities, such deviations can compromise the robustness of the sum rules.
As indicated by Figs.~\ref{fig:4mode_int_ge} and \ref{fig:3mode_int_ge}, interference with the vector operator can affect the deviations. 
This underscores the need for improved determinations of all relevant form factors for transitions into excited hadrons in order to achieve robust predictions based on the sum rules.

\bibliographystyle{utphys28mod}
\bibliography{ref}
\end{document}